\documentclass[final,5p,times,twocolumn]{elsarticle}
\usepackage{tikz}
\newcommand\copyrighttext{%
  \footnotesize \copyright 2019. This manuscript version is made available under the CC BY 4.0 International License http://creativecommons.org. The published journal article is available at: \href{https://www.sciencedirect.com/science/article/abs/pii/S0306261919318483}{https://www.sciencedirect.com/science/article/abs/pii/S0306261919318483};\\ \href{https://doi.org/10.1016/j.apenergy.2019.114161}{https://doi.org/10.1016/j.apenergy.2019.114161}}
\newcommand\copyrightnotice{%
\begin{tikzpicture}[remember picture,overlay]
\node[anchor=south,yshift=10pt] at (current page.south) {\parbox{\dimexpr\textwidth-\fboxsep-\fboxrule\relax}{\copyrighttext}};;
\end{tikzpicture}%
}

\usepackage{graphicx}
\usepackage{amssymb}

\usepackage{lineno}

\usepackage{amsmath}

\usepackage{makecell} 

\usepackage{floatrow} 
\newfloatcommand{capbtabbox}{table}[][\FBwidth]

\usepackage{capt-of}

\usepackage[colorinlistoftodos]{todonotes}  

\usepackage{url}
\usepackage{hyperref}

\usepackage{tabularx}

\usepackage{float} 
\usepackage[caption = false]{subfig}

\newcommand\footnoteref[1]{\protected@xdef\@thefnmark{\ref{#1}}\@footnotemark}

\usepackage{booktabs} 

\usepackage{rotating} 

\usepackage{framed}
\usepackage{nomencl} 
\makenomenclature

\newcommand{\nomunit}[1]{%
\renewcommand{\nomentryend}{\hspace*{\fill}#1}}

\usepackage{etoolbox}
\renewcommand\nomgroup[1]{%
  \item[\bfseries
  \ifstrequal{#1}{A}{Selected abbreviations}{%
  \ifstrequal{#1}{V}{Variables}{%
  \ifstrequal{#1}{P}{Parameters}{}}}%
]}

\usepackage{changes}
\definechangesauthor[name={Wilko}, color=blue]{WH}
\definechangesauthor[name={Wided}, color=red]{WM}

\journal{Applied Energy}

\begin{document}

\begin{frontmatter}


\title{Regionalised heat demand and power-to-heat capacities in Germany \textendash~An open data set for assessing renewable energy integration}



\author[label1]{Wilko Heitkoetter}\author[label1]{Wided Medjroubi}\author[label1]{Thomas Vogt}\author[label1]{Carsten Agert} 
\address[label1]{DLR Institute of Networked Energy Systems, Carl-von-Ossietzky-Str. 15, Oldenburg, Germany, wilko.heitkoetter@dlr.de}

\begin{abstract}
Higher shares of fluctuating generation from renewable energy sources (RES) in the power system lead to an increase of grid balancing demand. One approach for avoiding RES curtailment, is to use excess electricity feed-in for heating applications. To assess in which regions power-to-heat~(PtH) technologies can contribute to RES integration, detailed data on the spatial distribution of the heat demand are needed. 
In this contribution, we determine the overall heat load in the residential building sector and the share covered by electric heating technologies for each administrative district (NUTS-3) in Germany, with a temporal resolution of 15 minutes. For the detailed regionalisation of the residential building stock, we ordered a special evaluation of the German census data at the Research Data Centre with a combination of six building features.
Using these data, 729 building categories were defined and heat demand values were assigned to each category. 
Furthermore, heating types and different classes of installed heating capacity were defined.
Our analysis showed that the share of small scale single-storey heatings and large scale central heatings is higher in cities, whereas there are more medium scale central heatings in rural areas. Both is caused by the different shares of single and multi-family houses in the respective regions.
To determine the electrically covered heat demand, we took into account heat pumps and resistive heating technologies.
All results, as well as the developed code are {\hyperref[app_supp_mat]{published}} along with this contribution under open source licenses and can thus also be used by other researchers for the assessment of power-to-heat for RES integration. 

\end{abstract}


\end{frontmatter}

\copyrightnotice
\section{Introduction}

The expansion of renewable energy sources~(RES) is considered as a main instrument to reduce global CO$_2$ emissions and mitigate climate change~\cite{sims2004renewable}. In Germany, the share of RES in the power production increased tenfold, from $3\%$ in 1990 up to $32~\%$ in 2016~\cite{bloess2018power,hake2015german}. In the heating and cooling sector, the RES share is still significantly low, accounting only for $13~\%$ in 2016~\cite{bauermann2016german}. One option to increase the RES share in the heating sector, is to use biomass. However, its potential is limited by the large areas required for growing energy crops and the available amount of utilisable residues~\cite{hoogwijk2005potential}. 

Another option to increase RES share in the heating sector, is to convert power, that has been fed into the grid by wind and solar power plants, to heat~\cite{schweiger2017potential,averfalk2017large,bottger2014potential,bottger2015control,nyborg2015heat}. The most used power-to-heat~(PtH) technologies are resistive heaters and heat pumps~\cite{bloess2018power}. Since resistive heaters directly convert electricity into heat, the resulting ratio of heat output per power input, called coefficient of performance~(COP), is approximately $COP=1$. Heat pumps turn thermal energy from the surrounding, e.g. the ambient air or the ground, to utilisable heat. They can reach values of $COP>3$, depending on several influence factors, e.g. the heat source temperature~\cite{bloess2018power}.
Due to this high COP value, the specific CO$_2$ emissions of heat supply from heat pumps can be lower than the one from conventional gas or oil heating systems, even if a significant share of fossil fuel based power plants contributes to the electricity generation.

In cases of excess feed-in from RES, the PtH technology can contribute to the reduction of curtailment and grid overload~\cite{zhang2016reducing,goransson2014linkages}. For this purpose, scheduled PtH load is shifted from time periods of low RES power generation to periods with high generation.  As the actual heat demand of the users does not change temporally, a thermal storage is required, which can thus be regarded as an equivalent electricity storage. In most cases, heating systems are equipped with heat storage to buffer peak demand. Also the building's massive structure has a significant storage capacity~\cite{henze2004evaluation,lizana2017advances}. Since the thermal storage is already existing, costs can be saved, e.g. in comparison to installing a new battery storage. 

\begin{table*}[h]
\caption{Overview of selected studies taking into account heat demand regionalisation and PtH potential determination}
\scriptsize
\begin{center}
\begin{tabularx}{\textwidth}{X|X X X X X X X}
\toprule
Study & Spatial scope &Spatial resolution & Regionalisation Method & Building details & Demand sectors: residential/\newline commercial/\newline industrial& Openness and\newline reproducibility& Centralised/\newline decentralised PtH potentials\\
\midrule
Gils~\cite{gils2012gis}& Europe & $0.5~km^2$& top-down & + &\checkmark / \checkmark/-& ++&- / -\\
Persson~\cite{persson_2017,connolly2014heat}& Europe & $0.01~km^2$& bottom-up&++ &\checkmark / \checkmark/-&+&\checkmark / \checkmark \\
Corradini~\cite{corradini_2012} & Germany & municipality& bottom-up& ++ &\checkmark / - / -& ++&- / -\\
Pellinger~\cite{ffe_merit_2016}& Germany & municipality& bottom-up& ++ &\checkmark / \checkmark / \checkmark& ++&\checkmark / \checkmark\\
present study& Germany & admin. district& bottom-up&+++ &\checkmark / - / -&+++& \checkmark / \checkmark\\
LUBW / IWU~\cite{lubw_2016_energieatlas}& German fed. state&admin. district& bottom-up &+++ &\checkmark / - / -&++&- / -\\
\bottomrule
\multicolumn{8}{l}{Annotation: - = not considered; \checkmark = considered; + = limited; ++ = medium; +++ = high}
\end{tabularx}%
\label{tab_literature_overview}
\end{center}
\end{table*}

Besides the storage capacity, there are other constraints which limit application of PtH. It can be distinguished between a theoretical potential, its subsets, the technical, economic and social potential, and the resulting practical potential~\cite{gils2015balancing,grein2011load}. In this contribution, only the least constrained, theoretical potential is regarded.
We henceforth denote the thermal heating capacity of electrically driven heating devices that is installed, respectively will be installed in future, as "PtH potential". 

To assess, how PtH can contribute to a successful integration of RES in Germany, it is not sufficient to calculate one PtH potential value for the entire country. The curtailment of renewable power plants due to a surplus feed-in and limited grid transport capacity is often a local phenomenon, depending on the installed capacity and the energy demand in the respective region, as well as on the power flow from or to other regions~\cite{nep2017nep}. Therefore, PtH potential data for every grid region are required.

A main driver for the PtH potential within a region, is the corresponding heat demand.
Throughout this paper we will refer to the process of splitting up the heat demand of a continent or country to its associated regions as "regionalisation". It can be distinguished between top-down and bottom-up regionalisation approaches. Also the number of building characteristics that are taken into account for the regionalisation varies between the different approaches.  

Another important driver of the PtH potential in a region is the share of the heat demand that is covered by PtH technologies. It can be distinguished between decentralised PtH in individual buildings and centralised PtH in district heating grids. Table~\ref{tab_literature_overview} shows an overview of selected studies considering either heat demand regionalisation or the determination of PtH potentials.

Gils~\cite{gils2012gis} developed a top-down approach for the regionalisation of the heat demand on a European scale, taking into account residential and commercial buildings. The demand data were extracted from national energy balances and scenarios. They were then allocated according to land use and population density using a raster with approx. $0.5~km^2$ pixel size. Further, it was assumed that multi-family buildings have a~$20\%$ lower heat demand than single-family buildings.

Persson et al.~\cite{persson_2017,connolly2014heat} account for more building characteristics, using a complex bottom-up approach to create the Pan-European Thermal Atlas. 
It comprises residential buildings as well as the service sector and has a resolution of 100x100m. The authors used input data from the Danish National Building Register, the census data, Corine land cover data and the European settlement map. 
The data were processed in a floor area regression model and a heat and cooling demand density model. Overall PtH potentials for Europe were determined, but the PtH technology was not in the focus of the study. 
The results can be visualised on an online map\footnote{\href{http://stratego-project.eu/pan-european-thermal-atlas/}{http://stratego-project.eu/pan-european-thermal-atlas/}} but the underlying data are not publicly available. Due to the high model complexity, the results are also hardly reproducible.

The Institute for Housing and Environment (IWU) developed a comprehensive topology of the German building stock~\cite{iwu2010_datenbasis,iwu2015wohngebaeudetypo}. 
The Baden-Wuerttemberg State Institute for the Environment (LUBW) used these data to regionalise the heat demand of the German State Baden-Wuerttemberg to its associated administrative districts~\cite{lubw_2016_energieatlas}, taking into account multiple building properties that influence the heat demand: the floor area, year of construction and building type. The resulting data set is publicly available\footnote{\href{http://udo.lubw.baden-wuerttemberg.de/projekte/pages/selector/index.xhtml;jsessionid=A968F0874077E7906BD5B3D9386551A3.projekte2}{For more information visit: https://www.lubw.baden-wuerttemberg.de}}, but no source code used for the modelling is provided.

Corradini et al.~\cite{corradini_2012} determined the residential heat demand for each municipality in Germany. The authors combined multiple statistics for the year of construction of the buildings, settlement type, number of flats per building and floor area. Due to the usage of different data sources for the respective attributes, assumptions needed to be made, e.g. that the distribution of the buildings on the year of construction is independent of the building type. The results of~\cite{corradini_2012} are further used in~\cite{ffe_merit_2016}, where a merit order of energy storage for the 2030 horizon is modelled. Also centralised and decentralised PtH as an equivalent energy storage are investigated. The results of~\cite{corradini_2012} and~\cite{ffe_merit_2016} are not provided as open data.

In the listed studies, data on regionalised heat demand and PtH potentials are determined and are to some extent suitable for the investigation of how PtH can contribute to avoid RES curtailment and grid congestion. However, the studies do not consider several aspects that are covered in this contribution, as described in the following paragraphs.

In contrast to the mentioned studies, we also provide all input data and results of this contribution as open data and publish the developed code as open source. The Open Source and Open Data approach has been chosen by the authors as it allows for an external evaluation of the models, the assumptions used as well as the obtained results~\cite{medjroubi2017open,morrison2018energy,pfenninger2018opening}. Moreover, publishing data and models allows their use by other researchers and helps reducing the duplication of data collection. 
Some of the listed studies already took into account more than one building characteristic to regionalise the residential building stock. Though input data from multiple statistics have been combined, which may lead to inaccuracies and hampers the reproducibility of the results. To avoid this, we ordered a special evaluation of the German census enumeration results at the Research Data Centre of the German Federal Statistical Office. The resulting data set contains a cross combination of six residential building attributes that influence the heat demand. By this cross combination, 729 building categories are defined, for which each the number of buildings per administrative district in Germany is given.

After assigning the heat demand to the regionalised building stock and calculating the electrically covered share of the heat load, the above listed studies provide aggregated demand values per region. To the extent of our knowledge, the present contribution is the first that further categorises the heat demand in a region according to the size of the installed heating capacity in the individual buildings. This categorisation was introduced due to the following reason.
Considering the case that buildings are heated by PtH technology and the scheduled load shall be shifted to use up surplus generation from RES. For this purpose the heating devices need to be equipped with information and communication technology (ICT), like controllers or smart meters. 
Both, a small scale single-storey heating, as well as a large scale central heating need to be equipped with such ICT devices. Thus the specific ICT costs for making the PtH load shiftable, would be higher for a small scale heating device than for a large scale device. Hence the categorisation of the installed heating capacity, provided in this contribution, is an important input for a subsequent economic assessment of the shiftable PtH operation.  

In particular, this contribution will examine the following research questions:
\begin{itemize}
\setlength\itemsep{0em}
\item What are the residential heat demand and PtH potentials at the administrative districts level in Germany? Which potentials can be expected in the future? 
\item What are the differences between the regions potentials? How are the heat demand and PtH potential broken down on different heating capacity size classes?
\item What are the effects of the regionalisation parameters on the obtained potentials?
\end{itemize}
As part of the region4FLEX\footnote{The region4FLEX model is developed at the DLR Institute of Networked Energy Systems. \href{https://wiki.openmod-initiative.org/wiki/Region4FLEX}{For more information visit: https://wiki.openmod-initiative.org/wiki/Region4FLEX}} model, the results of this study will be enhanced by the identification of heat storage potentials and the modelling of the demand shifting process. Further, region4FLEX will allow for determining the regionalised flexibility demand of the German power grid as well as the economic assessment of PtH and other flexibility options.

The remaining part of this contribution proceeds as follows:
Section~\ref{sec_methods} describes the methods and data used to derive the regionalised heat demand and PtH potential data. In Section~\ref{sec_results} the results are presented and validated against literature values. We highlight the main conclusions of the study and give an outlook in Section~\ref{sec_conclusion_outlook}.  

\section{Methods}\label{sec_methods}
\begin{figure}[h!]
\centering\includegraphics[width=\linewidth]{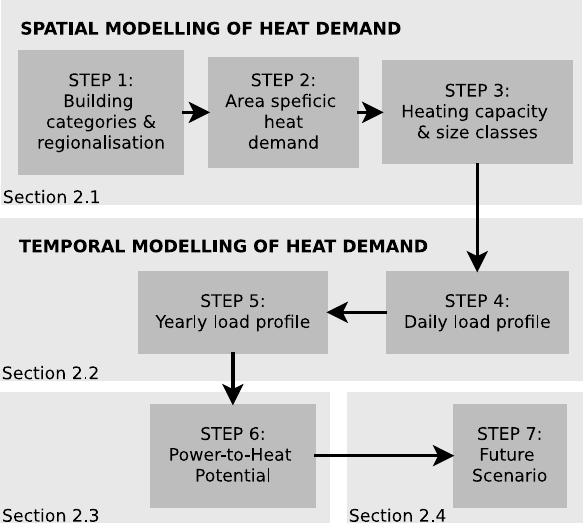}
\caption{Overview of the process for deriving the regionalised heat demand and power-to-heat potentials}
\label{methods_process_overview_SHORT}
\end{figure}

In this section, our seven-step approach for determining the regionalised heat demand and PtH potentials is introduced. As shown in Figure~\ref{methods_process_overview_SHORT}, the process consists of the spatial and temporal heat demand modelling, as well as the determination of the electrically covered heat load (PtH potential) and the derivation of a future scenario. The implementation of the process is described in~\ref{app_implementation}. 

\begin{table}[!t]   
\begin{framed}
\small
\nomenclature[A]{A/V}{Surface area to volume ratio}
\nomenclature[A]{DHW}{Domestic hot water}
\nomenclature[A]{SH}{Space heating}

\nomenclature[V,01]{$A$}{Floor area \nomunit{$m^2$}}
\nomenclature[V,02]{$n_{res}$}{Number of residents\nomunit{$-$}}
\nomenclature[V,03]{$T_a$}{Ambient temperature \nomunit{$^\circ C$}}
\nomenclature[V,04]{$T_{hl}$}{Heating limit temperature \nomunit{$^\circ C$}}

\nomenclature[V,05]{$Q$}{Annual heat demand \nomunit{$kWh/a$}}
\nomenclature[V,06]{$Q_{DHW}$}{Annual heat demand for\\ domestic hot water\nomunit{$kWh/a$}}
\nomenclature[V,07]{$Q_{SH}$}{Annual heat demand for space heating\nomunit{$kWh/a$}}

\nomenclature[V,08]{$q''$}{Annual area specific heat demand \nomunit{$kWh/m^2/a$}}
\nomenclature[V,09]{$q_{DHW}$}{Annual domestic hot water\\ demand per resident \nomunit{$kWh/cap/a$}}

\nomenclature[V,10]{$\dot{Q}_{inst}$}{Installed heating capacity \nomunit{$kW$}}
\nomenclature[V,11]{$\dot{Q}~(T_a)$}{Heat load \nomunit{$kW$}}
\nomenclature[V,12]{$t_{flh}$}{Full load hours\nomunit{$h$}}
\nomenclature[V,13]{$h$}{temperature and area specific \\heat load \nomunit{$kW/K/m^2$}}
\nomenclature[V,14]{$n_{hdd}$}{Heating degree days \nomunit{$K\cdot d$}}

\printnomenclature
\end{framed}
\end{table}

\subsection{Spatial modelling of the residential heat demand}
The spatial modelling process of the heat demand can be summarised as follows. First, we defined building categories and obtained the number of buildings per category for each administrative district from a special evaluation of the German census data. Then, for each building category, the respective heat demand was assigned, yielding also the overall heat demand per administrative district. Finally, we divided the buildings into classes according to the size of the required heating capacity. In the next paragraphs, these three steps of the spatial heat demand modelling will be elaborated.

\subsubsection*{STEP 1: Definition of building categories and regionalisation}\label{sec_meth_spatial_mod}

In the first step of the spatial modelling, residential building categories were defined and the number of buildings per category was assigned to each administrative district. We used data from the census enumeration, to model the German building stock.
Among others, the following attributes\footnote{A comprehensive description of the surveyed attributes is provided in~\cite{destatis2014overview}.} are surveyed in the census enumeration, that influence the thermal energy demand: type of building, number of flats per building, year of construction, floor area, heating type and number of residents per building. For each attribute multiple types are regarded. There are e.g. the types "detached", "semi-detached" or "row building" for the attribute "building type". 

The census enumeration data can be obtained from the website of the Federal Statistical Office of Germany\footnote{\label{note_census_website}\href{https://www.zensus2011.de/DE/Home/home_node.html}{For more information visit https://www.zensus2011.de}}. In the standard result tables on the website, it is listed, how many buildings per region (e.g. administrative district or municipality) have a specific type of an attribute. The website also contains a basic database query tool, in which up to three attributes can be combined. Thus the number of buildings in a selected region having a combination of specific types of attributes can be obtained. As the regionalisation of the building stock is in the focus of the present contribution, a combination of more than three building attributes was preferable. For this reason, we ordered a special evaluation of the census data~\cite{census_sonderauswertung} at the Research Data Centre of the German Federal Statistical Office and Statistical Offices of the L{\"a}nder\footnote{\href{https://www.forschungsdatenzentrum.de/en}{For more information visit https://www.forschungsdatenzentrum.de}} with a combination  of all six above mentioned building features that influence the heat demand.
\begin{figure}[h!]
\centering\includegraphics[width=0.75\linewidth]{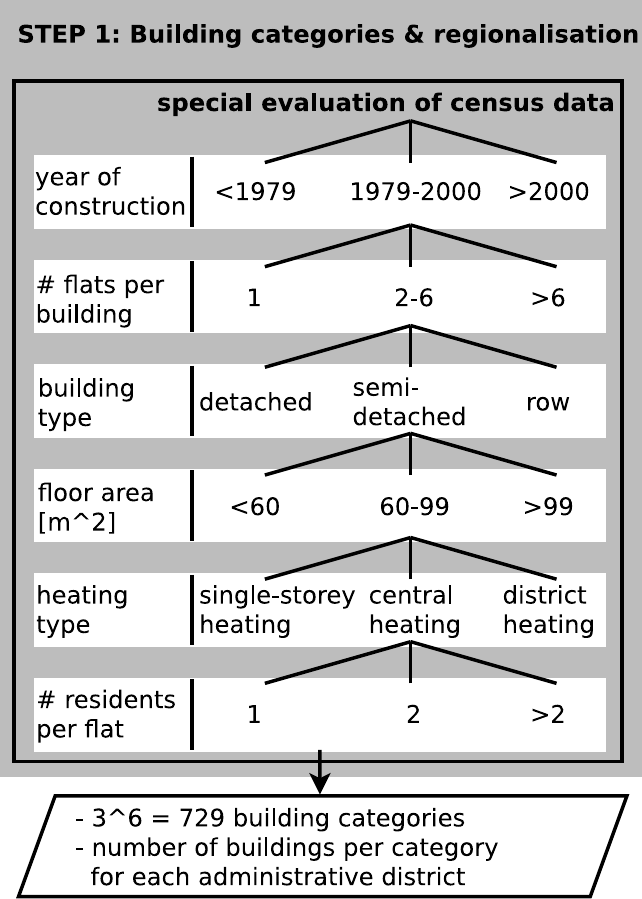}
\caption{The cross-combination of six building characteristics obtained from a special census evaluation.}
\label{fig_zensus_sonderauswertung_overview}
\end{figure}
Figure~\ref{fig_zensus_sonderauswertung_overview} shows the combination of these six attributes, considering three possible types for each attribute. For example the attribute "year of construction" has the types "$<1979$", "$1979-2000$" and "$>2000$". Each of the $3^6=729$ possible combinations represents one building category. In the result table of the special evaluation of the census data, the number of buildings per building category is given for all German administrative districts. The table can be found in the supplementary material section of this contribution\footnote{\href{https://doi.org/10.5281/zenodo.2650200}{Supplementary material/census special evaluation data}}.

The census data are also available with a higher spatial resolution than the administrative district level, e.g. on the municipality level or even on a $100x100m$ grid. But before publishing the data on the official website\textsuperscript{\ref{note_census_website}} or as part of a special evaluation,
the German Federal Statistical Office applies a security algorithm to the data. The algorithm, called SAFE~\cite{hoehne2015safe}, is applied due to privacy reasons. SAFE changes individual data, so that no information about individual persons can be gained. The changes are conducted in a way, that they compensate each other and the results for one geographic unit of the regarded data set (e.g. one specific administrative district) are changed as few as possible. The higher the spatial resolution of the regarded data set and the more evaluation attributes are combined, the higher is the relative modification of the data by the SAFE algorithm. In consultation with the Research Data Centre personal, we thus considered the spatial resolution on administrative district level as the best trade-off between data quality and the necessary spatial resolution for the aims of this study.



\subsubsection*{STEP 2: Area specific annual heat demand of the building categories}\label{sec_en_from_dena}

\begin{figure}[h!]
\centering\includegraphics[width=\linewidth]{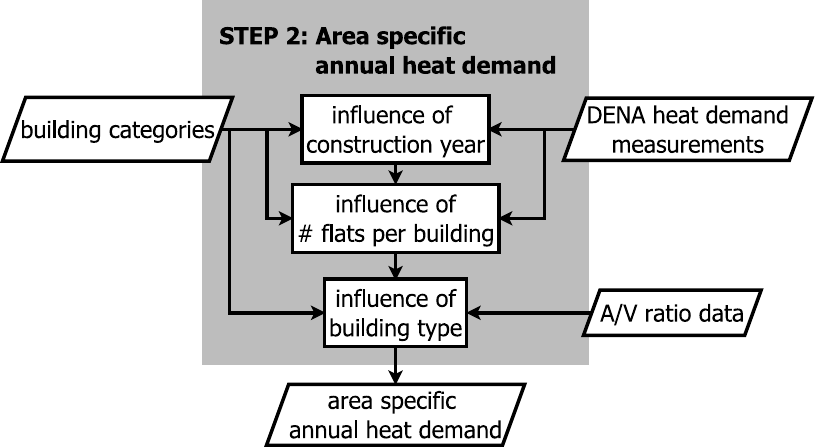}
\caption{Overview of the process to determine the floor area specific annual heat demand.}
\label{fig_step_2_area_spec_heat_demand}
\end{figure}

In the second step of the spatial modelling, we determined the floor area specific heat demand of the different building categories. As shown in Figure~\ref{fig_step_2_area_spec_heat_demand}, the influence of the attributes, year of construction, number of flats per building and building type, was taken into account.
The utilised input data were the 729 building categories, defined in the prior step, heat demand measurement data and data regarding the surface area to volume ratio ($A/V$~ratio) of buildings.

Several standards specify, how to calculate the area specific heat demand of individual residential buildings and how to design the respective heating system~\cite{din2011_18599,din2008_12831,vdi2008_3807}. These approaches require detailed information, like window surface areas or wall heat transfer coefficients. However, such data are not available when analysing a significant number of buildings, as e.g. all residential buildings in an administrative district. The building topology developed by the Institute for Housing and Environment (IWU)~\cite{iwu2010_datenbasis,iwu2015wohngebaeudetypo} is well suited to calculate the heat demand of the overall German residential building stock.
The IWU topology building features are not the same as in the special evaluation of the census data. 
Therefore we used other data sources that describe the direct dependency of the heat demand on the building attributes used in the special evaluation of the census data. 

To model the influence of the building attributes "year of construction" and "number of flats per building" on the annual area specific heat demand~$q''$, we used data of the German Energy Agency (DENA)~\cite{bigalke2015dena_geb_rep} that are based on measurements.
\begin{figure}[ht]
\centering\includegraphics[width=\linewidth]{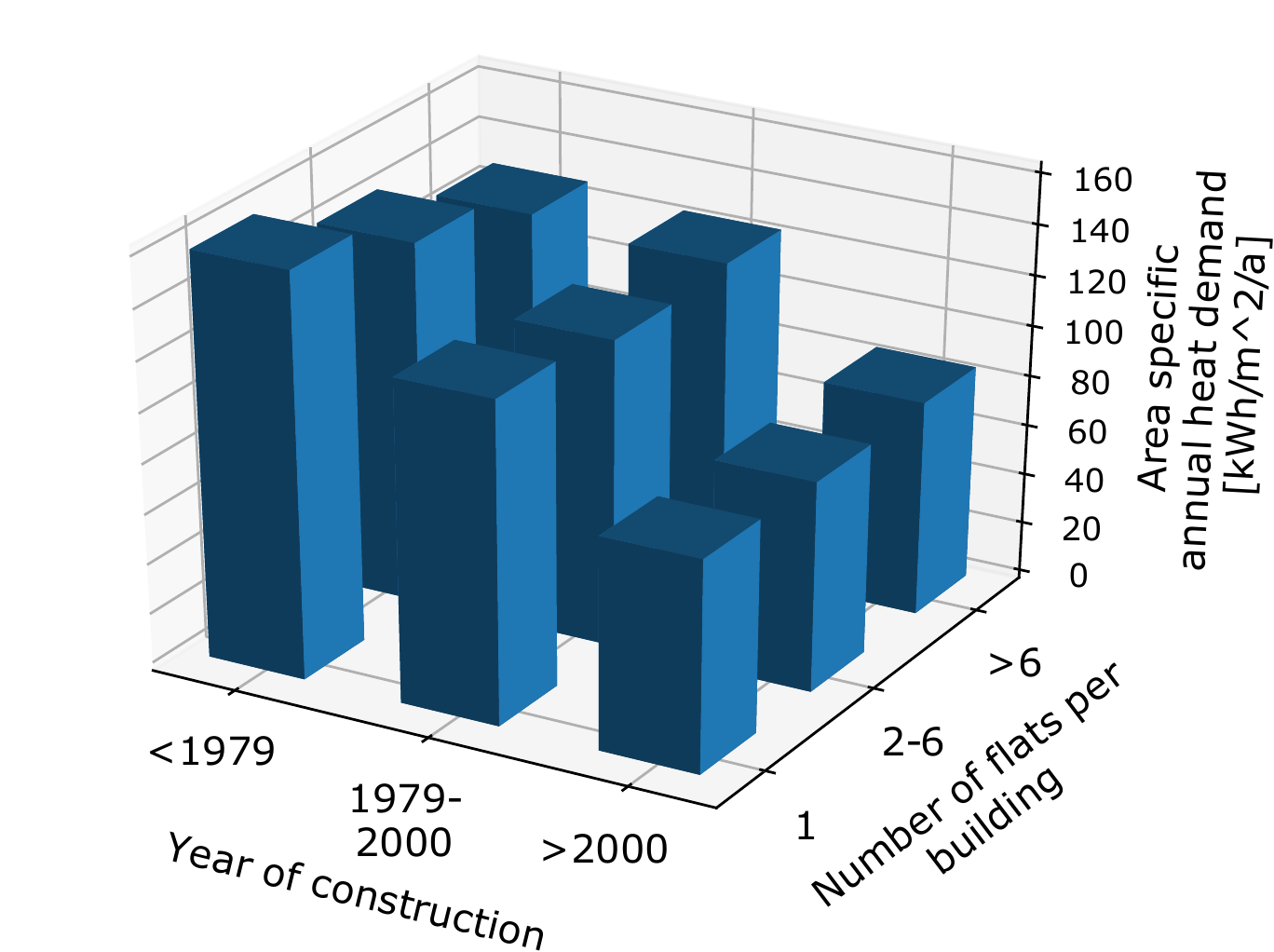}
\caption{Annual floor area specific final energy consumption for space heating and DHW, following~\cite{bigalke2015dena_geb_rep}.}
\label{fig_heat_dem_dena}
\end{figure}
The DENA values comprise the final energy consumption for both space heating and domestic hot water~(DHW). In Figure~\ref{fig_heat_dem_dena}, the data are visualised and the numerical values are given in the supplementary material of this contribution\footnote{\href{https://doi.org/10.5281/zenodo.2650200}{Supplementary material/other input data/heat demand according to year of construction and number flats per building.xls}}. The area specific heat demand of the buildings with the oldest year of construction ($<1979$) is about two times the demand of the buildings with the newest year of construction ($>2000$), which is caused by the different quality of insulation used.
For the buildings built before $1979$, the energy demand drops significantly with higher number of flats per building. The reason for this is that buildings with multiple flats have a lower ratio of surface area to building volume and thus a relatively lower area for heat transfer. 
For newer buildings, the number of flats has a lower influence on the heat demand, due to a more efficient insulation. 

\begin{figure}[ht]
\centering\includegraphics[width=\linewidth]{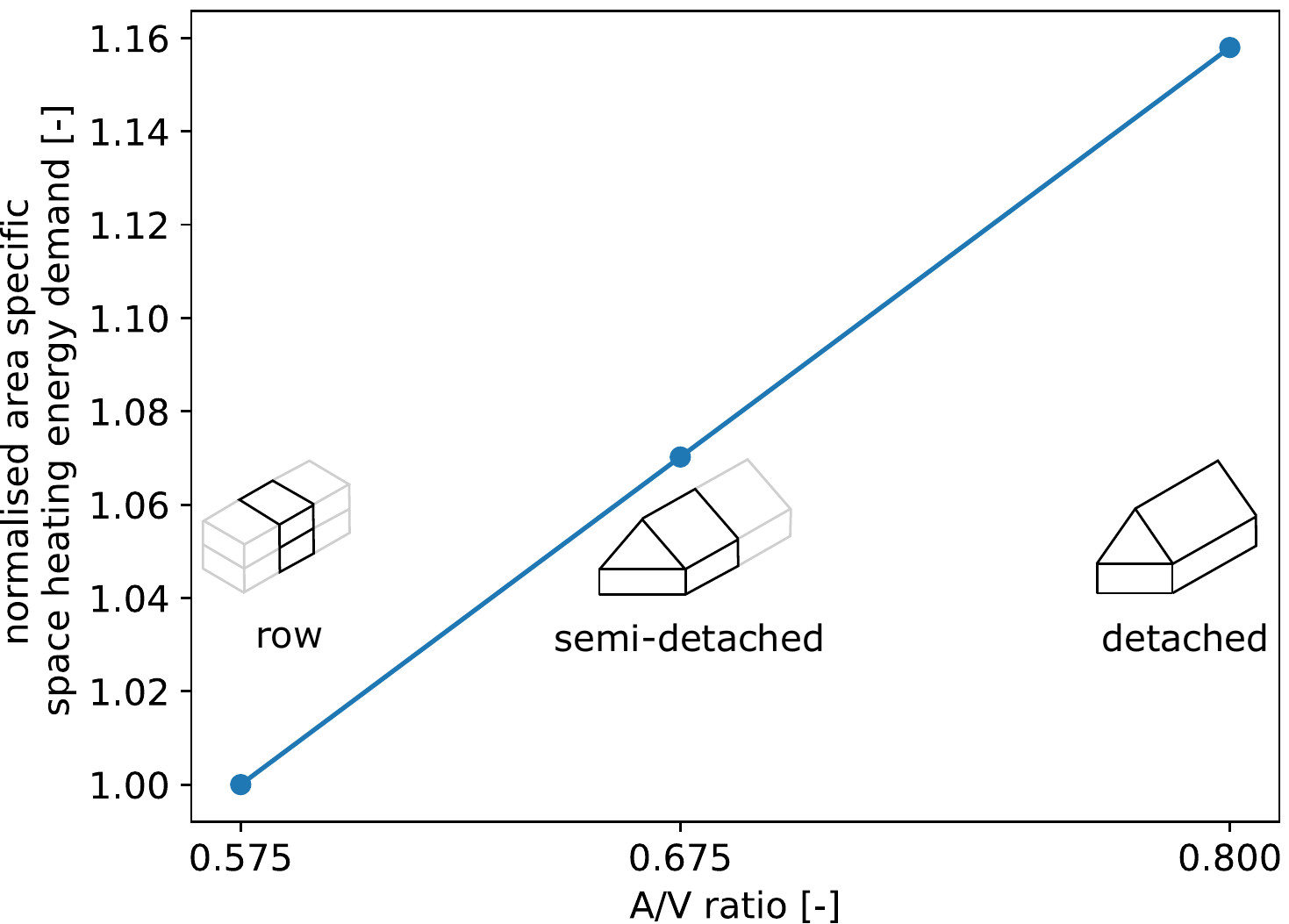}
\caption{A/V ratio of building types~\cite{glombik2008energieeffiziente} and influence on heat demand, normalised by the heat demand of the row building type.}
\label{fig_av_ratio_build_typ_heat_dem}
\end{figure}

Also the attribute "building type" influences the A/V ratio and thus the area specific heat demand. In Figure~\ref{fig_av_ratio_build_typ_heat_dem}, the three considered building types are assigned to their respective A/V ratio, using data provided in~\cite{glombik2008energieeffiziente}.
The influence of the A/V ratio on the heat demand can be calculated by Eq.~\ref{eq:heat_dem_a_v_dep}, which is given in~\cite{holtfort2002enev}:
\begin{equation} \label{eq:heat_dem_a_v_dep}
q''(A/V) = 50,94 + 75,29 \cdot A/V + 2600/(100+A).
\end{equation}
\noindent Applying the equation yields the results shown in Figure~\ref{fig_av_ratio_build_typ_heat_dem}, using an exemplary floor area of $A=150~m^2$. 
By this procedure we disaggregated the heat energy demand on the different building categories defined in Section~\ref{sec_meth_spatial_mod}, depending on the building type. More details are given in~\ref{app_influ_build_type}.



\subsubsection*{STEP 3: Absolute heat demand, heating types and capacity classes}

\begin{figure}[ht]
\centering\includegraphics[width=\linewidth]{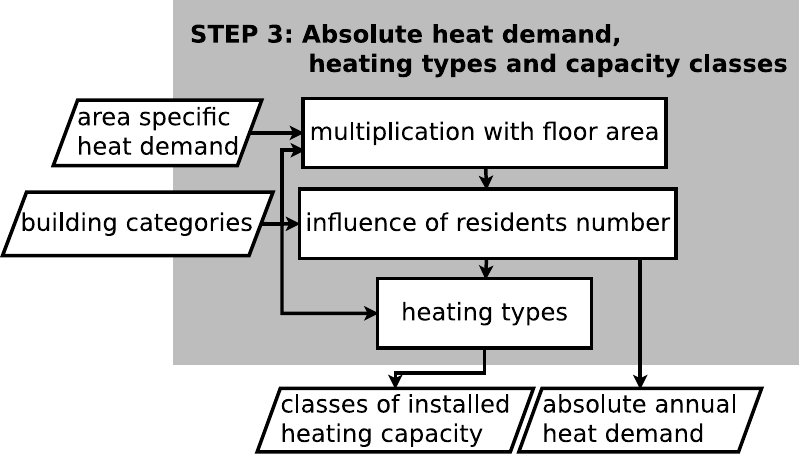}
\caption{Overview of the process to derive the absolute heat demand and the classes of installed heating capacity.}
\label{fig_meth_step_overview_abs_load_capacity_classes}
\end{figure}

In the third step of the spatial modelling, we calculated the absolute annual heat demand of the building categories, as well as the overall heat demand of the administrative districts, and introduced classes of installed heating capacity. Figure~\ref{fig_meth_step_overview_abs_load_capacity_classes} shows the input data and the performed substeps, which are further elaborated in the following paragraphs. 
To get the absolute annual heat demand $Q$ of the building categories, we multiplied the area specific heat demand, that was obtained in the prior step, with the average floor area, $Q=q''\cdot A$. The used floor area values are presented in the supplementary material\footnote{\href{https://doi.org/10.5281/zenodo.2650200}{Supplementary material/other input data/floor area.xls}}. 

Subsequently, we regarded the influence of the attribute "number of residents per flat" on the DHW demand. In the data from DENA~\cite{bigalke2015dena_geb_rep} that were used to calculate the area specific heat demand in the prior step, the energy demand for both space heating and DHW heating was aggregated. To disaggregate these two energy demand types, we derived an average annual DHW heating demand per resident. Therefore we divided the overall final energy demand for DHW in Germany, $90.12~TWH$ in 2010~\cite{bmwi2018energiedaten}, by the number of inhabitants in Germany, $80.21\cdot 10^6$~inhabitants according to the census 2011~\cite{destatis2011census}, which yielded $q_{DHW}=1123~kWh/cap/a$. This value was then multiplied with the average number of residents per flat and the number of flats per building. The resulting absolute DHW demand per building~$Q_{DHW}$ was subtracted from the overall heat demand, yielding the space heating demand of the building, $Q_{SH}=Q-Q_{DHW}$. For more details, refer to the supplementary material of this contribution\footnote{\label{foot_note_num_res}\href{https://doi.org/10.5281/zenodo.2650200}{Supplementary material/other input data/number of residents per flat.xls}}.     

From the absolute annual heat demand, also the installed heating capacity $\dot{Q}_{inst}$ could be approximated:
\begin{equation} \label{eq:heat_pow}
\dot{Q}_{inst} = \frac{Q}{t_{flh}},
\end{equation}
\noindent where $t_{flh}$ are the full load hours and following~\cite{Nabe_Prognos_2011}, we assumed $1900~h$ as full load hours for heating systems. Next, we took into account the attribute "heating type". We grouped the buildings according to the three types, single-storey heating, central heating and district heating.
For the central heating technology, it was further distinguished between
three capacity classes, based on the size of the heating, $\dot{Q}_{inst}<12.5~kW_{th}, 12.5~kW_{th}<\dot{Q}_{inst}<25~kW_{th}$ and $25~kW_{th}<\dot{Q}_{inst}$. For district heating, we summed the heating load of all buildings in each district having this heating type. The steps one, two and three of the spatial modelling, result in the determination of the installed heating load for each administrative district in Germany. 
In order to use the results of the present study for investigating the coupling of the heat sector with the power sector, the data on administrative district level may be assigned to the respective electricity grid districts.

\subsection{Temporal modelling of the residential heat demand}
In this section, the temporal modelling of the heat demand is described, which comprises step four and five of the overall modelling process. 
In step four, we calculated the daily load profile of the heat demand for different weekdays with a resolution of 15 minutes. These intraday load profiles are dominated by the usage patterns of the residents and the control settings of the heating system.
In step five, the time series of the heat demand in the course of the year was regarded. First, a yearly load profile with a resolution of one day was determined, which is dominated by the daily average ambient temperature. Then, the daily average load values were multiplied with the  normalised intraday load profiles that were calculated in step four, thus yielding a yearly load profile with a resolution of 15 minutes.  

\subsubsection*{STEP 4: Determination of daily load profiles}\label{sec_meth_intraday_load_profile}
For modelling the intraday resolution of the heat load, we used measurements that have been carried out by the DLR Institute of Networked Energy Systems within the NOVAREF project~\cite{lange_2018_novaref}. In the NOVAREF project, the thermal load for space heating and domestic hot water was measured for 12 residential buildings. The measurements were performed for at least one year for each building, with a time resolution of two seconds. Note that, in order to design a heating system for a single building, the high temporal resolution used in the NOVAREF project is beneficial, since it comprises demand peaks and ramps. 
Instead, the scope of the present contribution is to analyse the statistical demand time series of multiple thousands of buildings, e.g. in one administrative district.
Because such aggregated load profiles are smoother than those of individual buildings, we required a lower time resolution than in the NOVAREF project.  
Furthermore the results of this contribution shall be used to investigate balancing of fluctuating RES generation. Therefore, we chose a time resolution value of 15 minutes for the load profiles, which is also a common time period in power trading~\cite{kiesel_2016_modelling}.

We thus aggregated the NOVAREF time series of each building from a two second resolution into 15-minute average values~${<}\dot{Q}_{15min}{>}$.
The resulting yearly time series of each building with a 15-minute resolution was then cut into 365 one-day-interval time series. Further, the daily load profiles were normalised by dividing the 15-minute average load values by the daily average load $|{<}\dot{Q}_{15min}{>}| = {<}\dot{Q}_{15min}{>}/{<}\dot{Q}_{24h}{>}$.
The normalised load profiles were grouped by working days, Saturdays and Sundays and for each of the obtained groups, all associated daily load profiles were averaged.

\subsubsection*{STEP 5: Determination of the yearly load profile}
\begin{figure}[ht]
\centering\includegraphics[width=\linewidth]{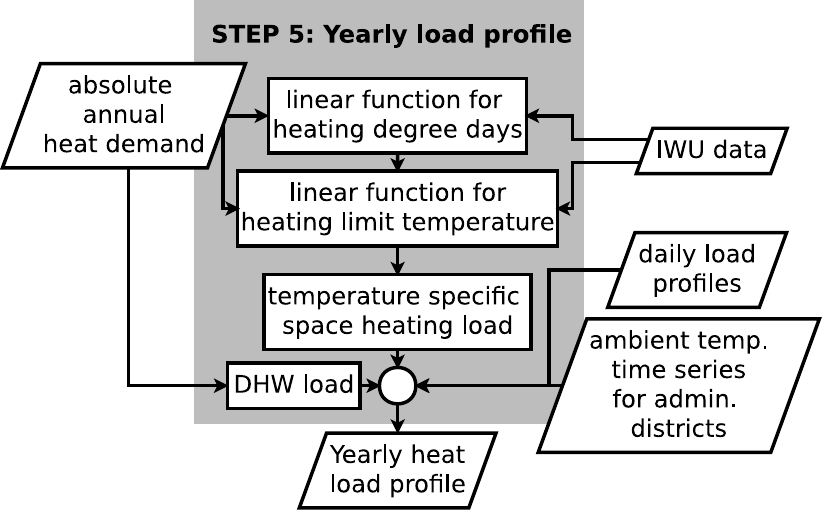}
\caption{Overview of the process for determining the yearly load profile.}
\label{diag_step_5_yearly_load_profile}
\end{figure}

In step five, we determined yearly load profiles for every considered building category and the aggregated heat load for each administrative district. An overview of this process is shown in Figure~\ref{diag_step_5_yearly_load_profile}.
First, the dependency of the space heating load on the ambient temperature was calculated. 
The resulting temperature specific space heating load was multiplied with the ambient temperature time series of the respective administrative districts, the building is located in.
Next, the DHW load was added to the space heating load and both were superposed with the intraday load profile.   
Finally, the aggregated heat load time series for each administrative district was calculated.
In the following paragraphs, this process is described in detail.

To model the dependency of the heat load $\dot{Q}(T_a)$ on the ambient temperature for each building category, we introduced a bilinear load profile, that is also used in several standards~\cite{din2011_18599,din2008_12831,vdi2008_3807}.
\begin{figure}[ht]
\centering\includegraphics[width=\linewidth]{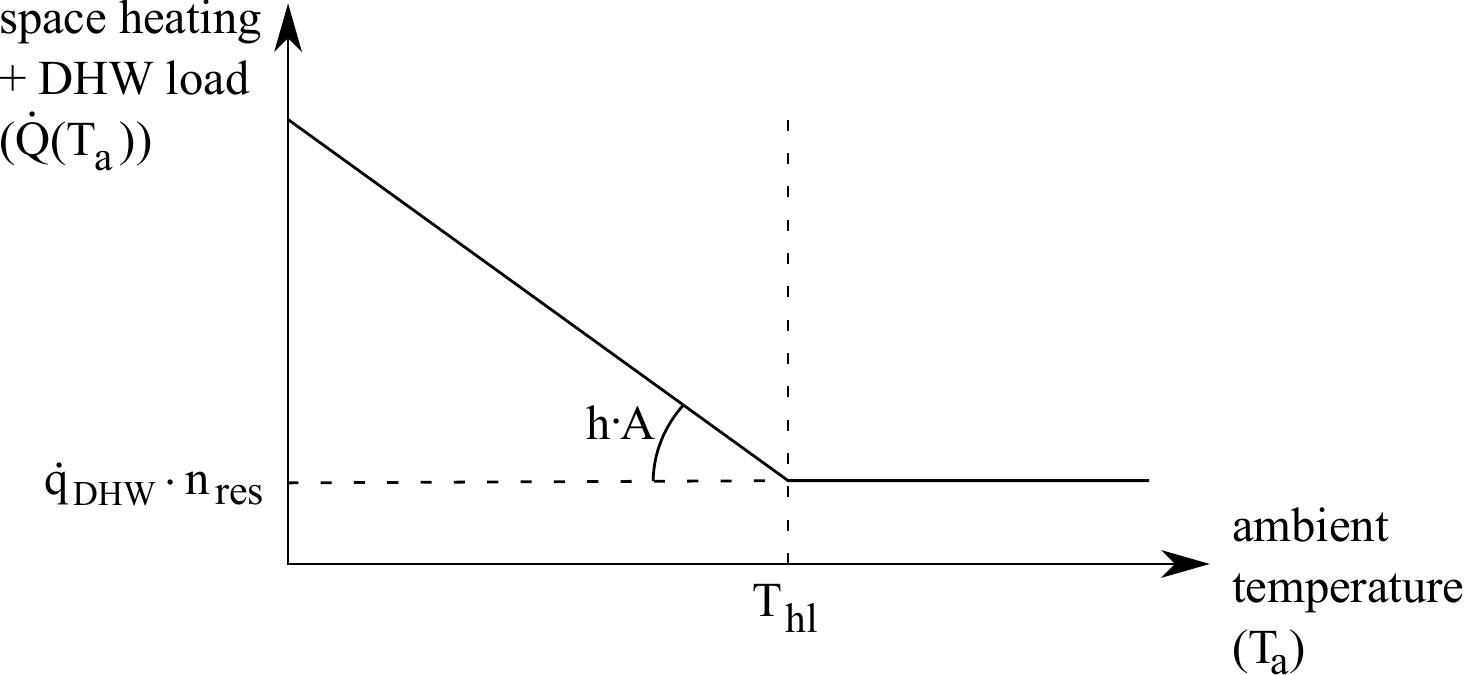}
\caption{Heat load dependency on ambient temperature.}
\label{diag_heat_load_t_ambient_schema}
\end{figure}
The profile is plotted in Figure~\ref{diag_heat_load_t_ambient_schema} and is defined by Eq.~\ref{eq_q_dot_ta_gr_thl} and~\ref{eq_q_dot_ta_sm_thl}: 
\begin{align}
T_a \geq T_{hl}: \dot{Q}(T_a) &= \dot{q}_{DHW}  \cdot  n_{res} \label{eq_q_dot_ta_gr_thl},\\
T_a < T_{hl}: \dot{Q}(T_a) &= h\cdot A\cdot (T_{hl} - T_a) + \dot{q}_{DHW} \cdot  n_{res}  \label{eq_q_dot_ta_sm_thl}.
\end{align}
When the ambient temperature $T_a$ is equal or larger than the heating limit temperature~$T_{hl}$, which depends on the attributes of the building, no space heating, just DHW heating is required. The DHW heat load per resident~$\dot{q}_{DHW}$ is assumed to be independent of the ambient temperature. Dividing the annual DHW heating demand per resident, introduced in Section~\ref{sec_meth_spatial_mod}, by the number of hours per year, $8760~h$, yields a value of $\dot{q}_{DHW}=0.128~kW_{th}$. Then $\dot{q}_{DHW}$ is multiplied with the number of residents per building $n_{res}$.  

In case of~$T_a < T_{hl}$, the space heating load rises linearly with the difference of $T_{hl}$ and $T_a$. The slope of the space heating curve is the temperature and area specific heat load $h$, multiplied with the floor area $A$. The equations~\ref{eq_q_dot_ta_gr_thl} and~\ref{eq_q_dot_ta_sm_thl} show that the heat load~$\dot{Q}(T_a)$ directly depends on the values of the building category attributes "number of residents per flat" and "floor area". The values of the specific heat load $h$ and heating limit temperature $T_{hl}$ are not defined, yet. We therefore derived functions for $h$ and $T_{hl}$ that are dependent on the annual heat demand for space heating $Q_{SH}$. Since we defined $Q_{SH}$
depending on the building category attributes, as described above, also $h$ and $T_{hl}$ are indirectly dependent on these attributes. 

We started the derivation of the functions for $h$ and $T_{hl}$ by defining $Q_{SH}$, not from measurement values as in Section~\ref{sec_meth_spatial_mod}, but by integrating the space heating load (first summand from Eq.~\ref{eq_q_dot_ta_sm_thl}) over time:
\begin{align} 
Q_{SH}&=\int_{t_{start}}^{t_{end}} h\cdot A\cdot (T_{hl} - T_a(t)) \ dt \label{eq_W_basic}\\
&=h\cdot A\cdot \int_{t_{start}}^{t_{end}} (T_{hl} - T_a(t)) \ dt \label{eq_W_ha_front}\\
&=h\cdot A\cdot n_{hdd} \label{eq_W_hdd}, \\
h&=\frac{Q_{SH}}{A\cdot n_{hdd}}\label{eq_h_W_A_nhdd},
\end{align}
where $t_{start}$ and $t_{end}$ are the time limits of the heating period. Since~$h$ and~$A$ are constants, they can be extracted from the integral (Eq.~\ref{eq_W_ha_front}). The remaining integral part yields the so called \emph{heating degree days}~$n_{hdd}$~[$K\cdot d$] (Eq.~\ref{eq_W_hdd})~\cite{din2011_18599,allen1976modified}, which in turn depend on the heating limit temperature and the location of the building. With an Excel tool\footnote{\href{t3.iwu.de/fileadmin/user_upload/dateien/energie/werkzeuge/Gradtagszahlen_Deutschland.xls}{For more information visit: https://iwu.de}}, developed by the Institute for Housing and Environment~\cite{iwu_exel_tool}, we calculated the number of heating degree days for three different heating limit temperatures, averaged over all weather stations in Germany. For each station, the average number of heating degree days of all years was used, that were measured at the respective station. The results are listed in Table~\ref{tab_hl_temp_n_hdd}.   
\begin{table}[ht]
\caption{Heating degree days for different heating limit temperatures~\cite{iwu_exel_tool} averaged over all weather stations in Germany.}
\centering
\begin{tabular}{p{0.25\textwidth} | p{0.25\textwidth}}
\hline
$T_{hl}~$[$^\circ$C] & $n_{hdd}$ [$K\cdot d$] \\
\hline
10 & 1254.78\\
12 & 1667.90\\
15 & 2409.75\\
\hline
\end{tabular}
  \label{tab_hl_temp_n_hdd}
\end{table}
We interpolated the average heating degree days values by using the following linear function:
\begin{equation}
n_{hdd}=-1087.3 +232.28\cdot T_{hl}. \label{eq_n_hdd_th_hl} 
\end{equation}

The heating limit temperature in Eq.~\ref{eq_n_hdd_th_hl} is not defined yet. According to~\cite{iwu_exel_tool}, the average heating limit temperature for low-energy houses is $12~^\circ C$ and for buildings with a weak heat insulation, it is $15~^\circ C$. Therefore, we assigned ${T_{hl}=12~^\circ C}$ to the building category (according to the definition in Section~\ref{sec_meth_spatial_mod}) with the lowest annual heat energy demand and ${T_{hl}=15~^\circ C}$ to the category with the highest heat energy demand. To determine the heating limit temperature values for the other building categories, we interpolated the outer limits by using a linear function, as follows: 
\begin{equation}
T_{hl}=10.46+0.0229\cdot \frac{Q_{SH}}{A}. \label{eq_t_hl_W_A} 
\end{equation}       
Inserting Eq.~\ref{eq_n_hdd_th_hl} and~\ref{eq_t_hl_W_A} into Eq.~\ref{eq_h_W_A_nhdd} yields:
\begin{align}\label{eq_h_full}
h&=\frac{Q_{SH}}{A\cdot 24\cdot (-1087.3 +232.28\cdot (10.46+0.0229\cdot \frac{Q_{SH}}{A}))}\\
&=0.00783324 - \frac{1.97679}{252.359\cdot \frac{Q_{SH}}{A}}.
\end{align}
A factor of 24 needed to be added in the denominator in order to convert days into hours. More details on the derivation of the equations~\ref{eq_n_hdd_th_hl}~-~\ref{eq_h_full} are provided in the supplementary material of this contribution\footnote{\href{https://doi.org/10.5281/zenodo.2650200}{Supplementary material/other input data/ heat load area and ambient temperature specific function.xls}}.  

\begin{figure}[h!]
\centering\includegraphics[width=\linewidth]{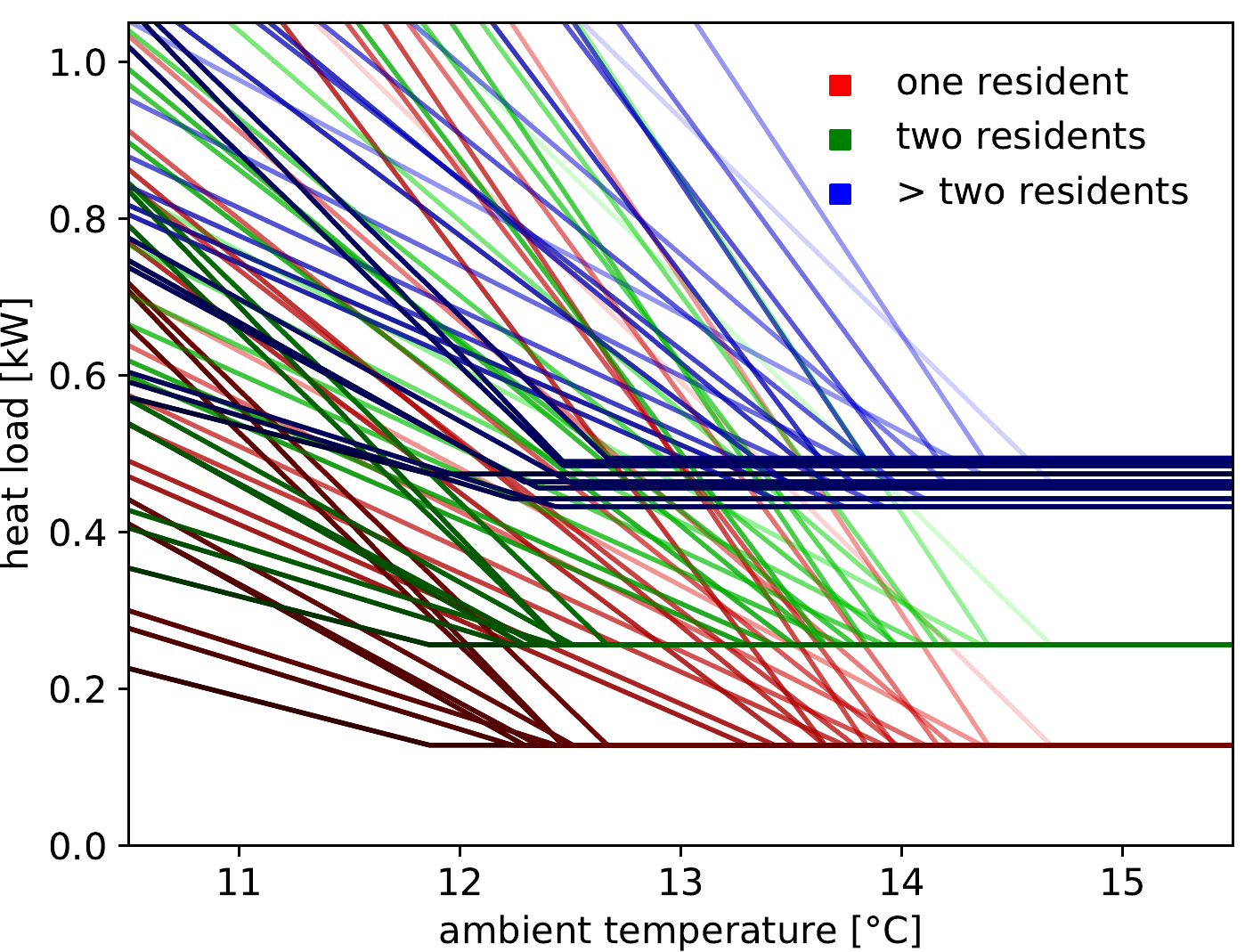}
\caption{Heat load of the different building categories in the heating limit temperature region. One line represents one building type. Dark colour: low heating limit temperature, bright colour: high heating limit temperature. Not all 729 load curves shown, for improved clarity.}\label{fig_Q_categories_T_hl_zoom}
\end{figure}

After deriving the functions for the heating limit temperature $T_{hl}$ and the specific heat load $h$ we evaluated Eq.~\ref{eq_q_dot_ta_gr_thl} and Eq.~\ref{eq_q_dot_ta_sm_thl} for all building categories that have been defined in Section~\ref{sec_meth_spatial_mod}. The resulting heat load over the ambient temperature is shown in Figure~\ref{fig_Q_categories_full}~(see~\ref{app_heat_load_curves}). Figure~\ref{fig_Q_categories_T_hl_zoom} (obtained when zooming Figure~\ref{fig_Q_categories_full}) shows the heat load of the different building categories with more details in the region of the heating limit temperature.
There are three groups of curves with a vertical offset between each other, which is caused by the different resident numbers in the building categories\textsuperscript{\ref{foot_note_num_res}}, influencing the DHW demand. The curves in Figure~\ref{fig_Q_categories_T_hl_zoom} have varying heating limit temperatures and different slopes, which is due to the difference in the year of construction, building type, floor area and number of flats per building.

To define the ambient temperature during the course of the year, we assigned the administrative districts to the respectively closest weather station. For each of the 44 weather station, the monthly average temperatures since the start of measurements were obtained from~\cite{iwu_exel_tool}\footnote{In~\cite{iwu_exel_tool} the temperature data were obtained from the \href{https://www.dwd.de/EN/Home/home_node.html}{German Meteorological Service}.}. The related data are provided in the supplementary material of this contribution\footnote{\href{https://doi.org/10.5281/zenodo.2650200}{Supplementary material/other input data/administrative districts closest weather station average temperatures.xls}}. 
We assigned the monthly average temperature to each fifteenth day of the month. The temperatures for the other days were linearly interpolated. Then in a next step, we inserted the average daily ambient temperature into Eq.~\ref{eq_q_dot_ta_sm_thl} for all building categories that have been defined in Section~\ref{sec_meth_spatial_mod}. This yielded the respective daily average space heating and DHW load. The average loads were then multiplied with the normalised intraday load profile with a 15 minute resolution for the respective day category (working days, Saturdays and Sundays).   

\subsection{Electrically covered heat load (STEP 6)}\label{sec_degree_electrification}
After modelling the spatial and temporal distribution of the heat demand in the previous steps, we continued by determining the PtH potential in step six. Therefore we calculated the share of the heat load, that is covered by an electric source.
 
We took into account the heat pump technology, for which we summed up the numbers of the air- and ground-sourced systems. Also we regarded the resistive space heating and resistive DHW heating technologies. For more information on the functionality of these technologies, refer to~\cite{bloess2018power}. Unfortunately, there are no data sets available, in which the number of installed electric heating systems are divided up into all of the census enumeration attributes, used to define the building categories in Section~\ref{sec_meth_spatial_mod}. But at least the two attributes "number of flats per building" and "year of construction" are considered.
Table~\ref{tab_heat_pump_distr} shows the number of buildings equipped with heat pumps~\cite{Nabe_Prognos_2011}, broken down according to the number of flats per building.
\begin{table}[h]
\caption{Number of installed heat pumps in Germany (2011) listed according to the number of flats per building~\cite{Nabe_Prognos_2011}.}
\small
\begin{center}
\begin{tabularx}{\textwidth}{ X|X }
\hline
category: flats per building & installed heat pumps per category [-] \\
\hline
1& 195000\\
2 - 6 & 123000\\
$> 6$ & 55000 \\
\hline
\end{tabularx}%
\label{tab_heat_pump_distr}
\end{center}
\end{table}

The number of flats with resistive space heating and resistive DHW heating systems are shown in Table~\ref{tab_resis_space_heat_DHW}, following~\cite{destatis_microcen_2010}, where the data are separated according to buildings with one flat and buildings with more than one flat. Also three different categories for the attribute year of construction of the building are distinguished.
\begin{table}[h]
\caption{Flats with resistive heating in Germany (2010) listed according to the number of flats per building and the year of construction~\cite{destatis_microcen_2010}.}
\small
\begin{center}
\begin{tabularx}{\textwidth}{ X|X|X X X }
\hline
PtH technology&category: flats per building&\multicolumn{3}{c}{number of flats with this PtH technology}\\
\hline
& &\multicolumn{3}{c}{year of construction of building}\\
& &1918-1979&1979-2000&$>2000$\\
\hline
resistive space &1& 223000 & 44000 & 8670 \\
heating &$>1$& 691000 & 119000 & 10340 \\
\hline
resistive DHW&1& 1296000 & 21300 & 46000 \\
heating& $>1$ & 4531000 & 707000 & 70000 \\
\hline
\end{tabularx}
\label{tab_resis_space_heat_DHW}
\end{center}
\end{table}
Further, we assumed that resistive space heating is only used in flats with single-storey heating.
Then the number of flats equipped with heat pumps and resistive heating technologies was divided by the total number of flats in each category. The resulting shares were multiplied with the totally installed heating capacity in the respective category, which yielded the installed electrically covered heat load. For more details on the numbers of flats equipped with PtH technologies, refer to the supplementary material of this contribution\footnote{\label{foot_note_el_heat_factors}\href{https://doi.org/10.5281/zenodo.2650200}{Supplementary material/other input data/electric heating factors}}.

The numbers listed in Table~\ref{tab_heat_pump_distr} and~\ref{tab_resis_space_heat_DHW} describe the share of the electrically covered heat load for entire Germany. In order to account for regional differences, we introduced a scaling factor for each German federal state. This factor specifies, whether in the respective federal state, the electrically covered share of the heat load is higher or lower than in the other federal states. The scaling factor of each federal state is multiplied with the electrically covered heat load share in all administrative districts associated with the particular state. More information on the calculation of the scaling factor are given in~\ref{app_PtH_state_scal_fac}.    

The thermal capacities of power-to-heat facilities in district heating systems in Germany are listed in~\cite{christidis_2017_eneff} (note~\ref{app_heat_stor_dist_heat} for heat storage in district heating systems). In this contribution, we assigned these capacities to the centroids of the respective administrative districts and plotted the resulting data in Figure~\ref{fig_heating_class_size_distribution_GER}. The numerical values are also listed in the supplementary material of this contribution\textsuperscript{\ref{foot_note_el_heat_factors}}.

\subsection{Future scenario derivation (STEP 7)}\label{sec_meth_fut_dev}
In the last step of the modelling process, we a applied a basic future scenario for the heat demand and the PtH potentials, taking into account the horizon 2030 and 2050.
We approximated that, in all German administrative districts, the relative increase or decrease of the heat load is equal. To describe the future development of the total thermal load of residential buildings, we used the values in~\cite{gils2015balancing}.

For decentralised PtH in residential buildings, we differentiated between the future projections of the installed load of heat pumps, resistive space heating and DHW boilers. To predict the installed thermal load of heat pumps in 2030, we used the values in~\cite{Nabe_Prognos_2011}. Furthermore we interpolated the values between the period 2010 and 2030 by a polynomial function of the second degree. The same function was used to extrapolate the installed heat pump capacities for the year 2050. The numerical values are given in the supplementary material\footnote{Supplementary material/other input data/ future projection}.

Resistive space heatings are mostly implemented as night storage heatings, using excess power from fossil or nuclear power plants~\cite{stadler2008gigantisches}. The energy transition and phase out of large scale power plants will probably lead to a reduction of the number of installed night storage heatings. At the same time, the need of balancing of renewable energy feed-in may lead to an increase in the numbers of resistive heaters~\cite{gassmann2018unerwartete}. We thus assumed that their overall load will remain constant. DHW boilers are mostly installed due to the low required space. Since this advantage will also be relevant in future, in this contribution it was assumed, that the overall load of this technology will too remain constant. 

For centralised PtH in district heating systems, no differentiation was undertaken between the future development of heat pumps and resistive heaters. This was mainly because of the lack of scenario data. In~\cite{energiewende2014power}, the future potential of PtH in district heating systems in Germany was assumed to be $4.5~GW$. The authors in~\cite{energiewende2014power} derived this number from the currently installed PtH capacity in Denmark. 
In this contribution, we assumed that these $4.5~GW$ will be installed by 2030. Furthermore the installed PtH capacity in district heating systems was assumed to grow by $50~\%$ from 2030 to 2050.   

\section{Results and Discussion}\label{sec_results}
In this section, we present the modelling results for the regionalised heat demand and PtH potential of the administrative districts in Germany. 
First, the findings for the installed heating capacity are described, considering different technologies and capacity sizes. Next, the temporal resolution and the future development of the heat load and PtH potentials are addressed. Finally, a validation of the results is conducted, for both the heat demand values of entire Germany and individual administrative districts.

\subsection{Installed heating capacity}

In the following subsections, we show the results for the modelling of the installed heating capacity in the German administrative districts. Differences between municipal and rural districts, as well as between different mayor cities, are especially highlighted. The electrically covered heat load and the shares of the different used technologies are also presented. In the supplementary material of this contribution\footnote{\href{https://doi.org/10.5281/zenodo.2650200}{Supplementary material/results}}, the numerical results for all German administrative districts can be found.

\subsubsection*{Structure of the installed heating capacity in municipal and rural districts}

\begin{figure}[ht]
\centering\includegraphics[width=0.85\linewidth]{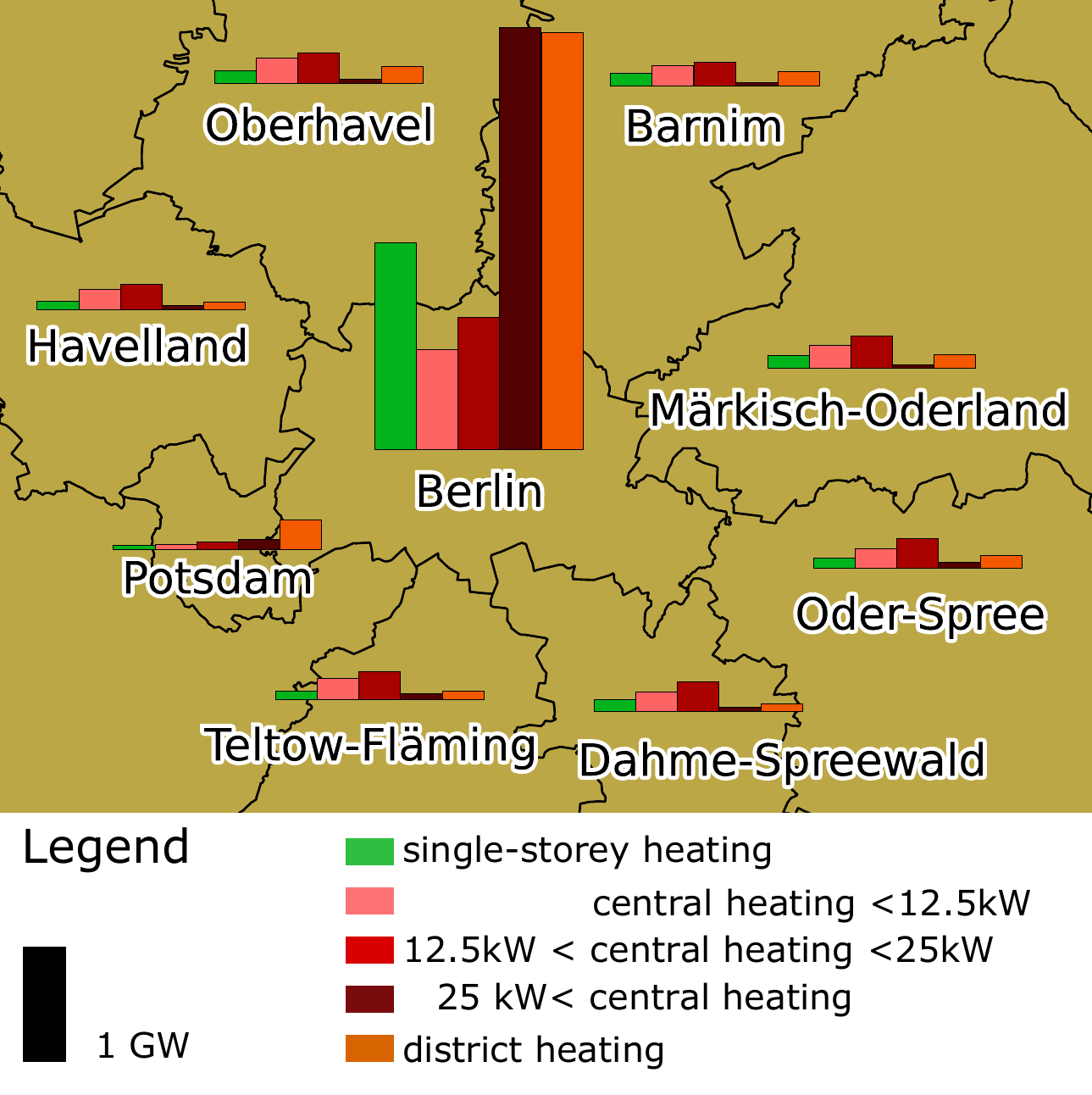}
\caption{Installed heating capacity and and heating technology distribution in the city of Berlin and the surrounding rural districts.}
\label{fig_map_berlin_region_heat_load}
\end{figure}

We begin with demonstrating the differences of the installed heating capacity and the heating technology distribution at the example of the city of Berlin ($population~density=4000~inhabitants/km^2$) and the surrounding rather rural\footnote{All administrative districts shown in Figure~\ref{fig_map_berlin_region_heat_load}, except Berlin and Potsdam, have a population density lower than $150~inhabitants/km^2$, which can be considered as rural, according to the definition in~\cite{oecd2011regional}.} administrative districts ($population~density<150~inhabitants/km^2$). As shown in Figure~\ref{fig_map_berlin_region_heat_load}, in Berlin, a high share of the residential heat load is covered by the district heating system. Single-storey heatings and large scale central heatings ($25~kW_{th}<\dot{Q}_{inst}$) also represent a large share of the heating technologies used. This is mainly the result of the high number of multi-family houses.
In the rural districts like Havelland, the highest share of the heat load is covered by small and medium scale central heating systems ($\dot{Q}_{inst}<25~kW_{th}$). The reason for this is that there are mostly single-family houses in the rural districts.

\begin{figure*}[p]
\centering\includegraphics[width=0.875\linewidth]{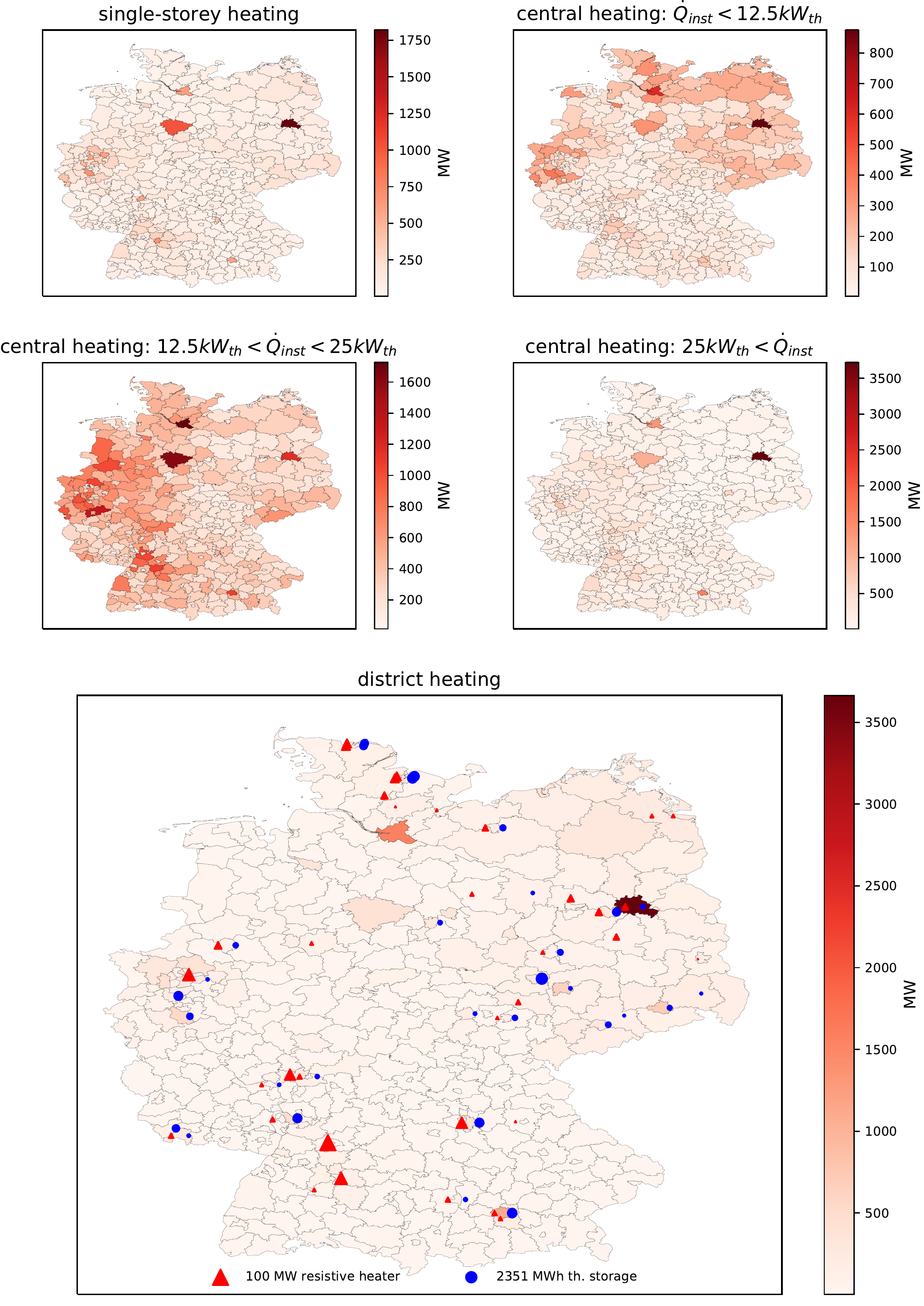}
\caption{Distribution of the installed heating capacity and centralised PtH in district heating systems in Germany.}
\label{fig_heating_class_size_distribution_GER}
\end{figure*}

The installed heating capacity for all districts in Germany, divided into the same size classes as for the case of the Berlin region, is shown in Figure~\ref{fig_heating_class_size_distribution_GER}. This yields that also for entire Germany, single-storey heating, large scale central heating and district heating are dominating in cities. Small and medium scale central heatings are dominating in rural areas.  


\subsubsection*{Structure of the installed heating capacity in different major cities}
\begin{figure}[h!]
\centering\includegraphics[width=\linewidth]{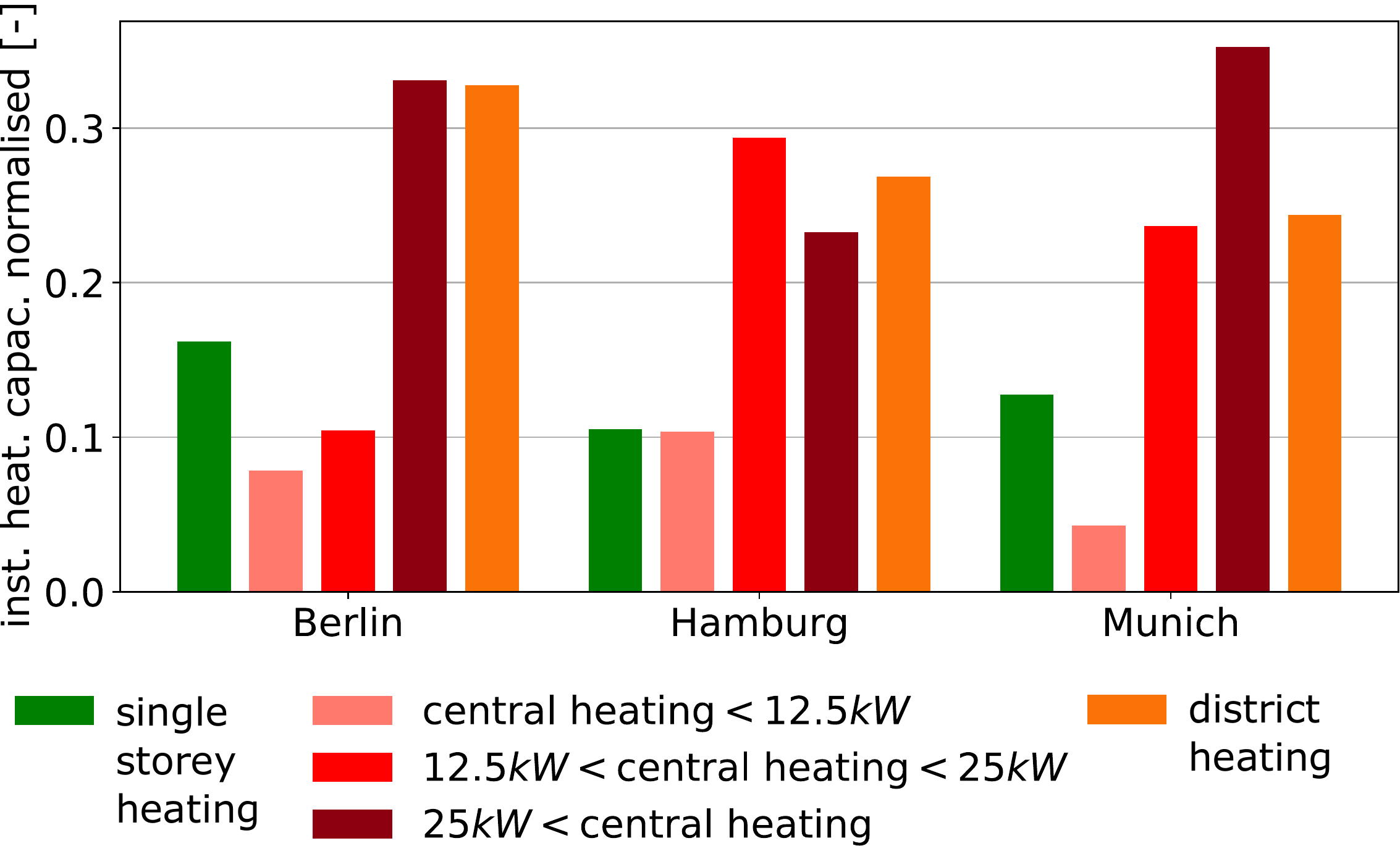}
\caption{Normalised distribution of the installed heating capacity in major German cities.}
\label{diag_diff_cities_heat_load}
\end{figure}

The distribution of the heat load in different major German cities, normalised by the totally installed heating capacity in the respective city, is shown in Figure~\ref{diag_diff_cities_heat_load}. 
Such as it was the case in Berlin, district heating and large scale central heating cover a significant percentage of the heat demand in the cities of Hamburg and Munich. The share of the medium scale central heating~($12.5~kW_{th}<\dot{Q}_{inst}<25~kW_{th}$) in Hamburg and Munich is more than twice as high as in Berlin. This may be explained by the lower number of single-family houses in Berlin due to a different housing policy in the former German Democratic Republic as in Western Germany~\cite{shlomo2011unterschiede}.

\subsubsection*{Share of electric heating technologies covering the heat load}
As described in Section~\ref{sec_degree_electrification}, we next multiplied the installed heating capacity with the electrically covered share, considering different technologies. 
This yields the theoretical PtH potential that can be used for load shifting in order to balance fluctuating renewable energy feed-in, which will be investigated by the authors in future research. 

\begin{figure}[h]
\subfloat[\label{diag_berlin_heat_load_electric_share}Totally installed heating capacity in comparison to the electrically covered share.]{\includegraphics[width=0.475\textwidth]{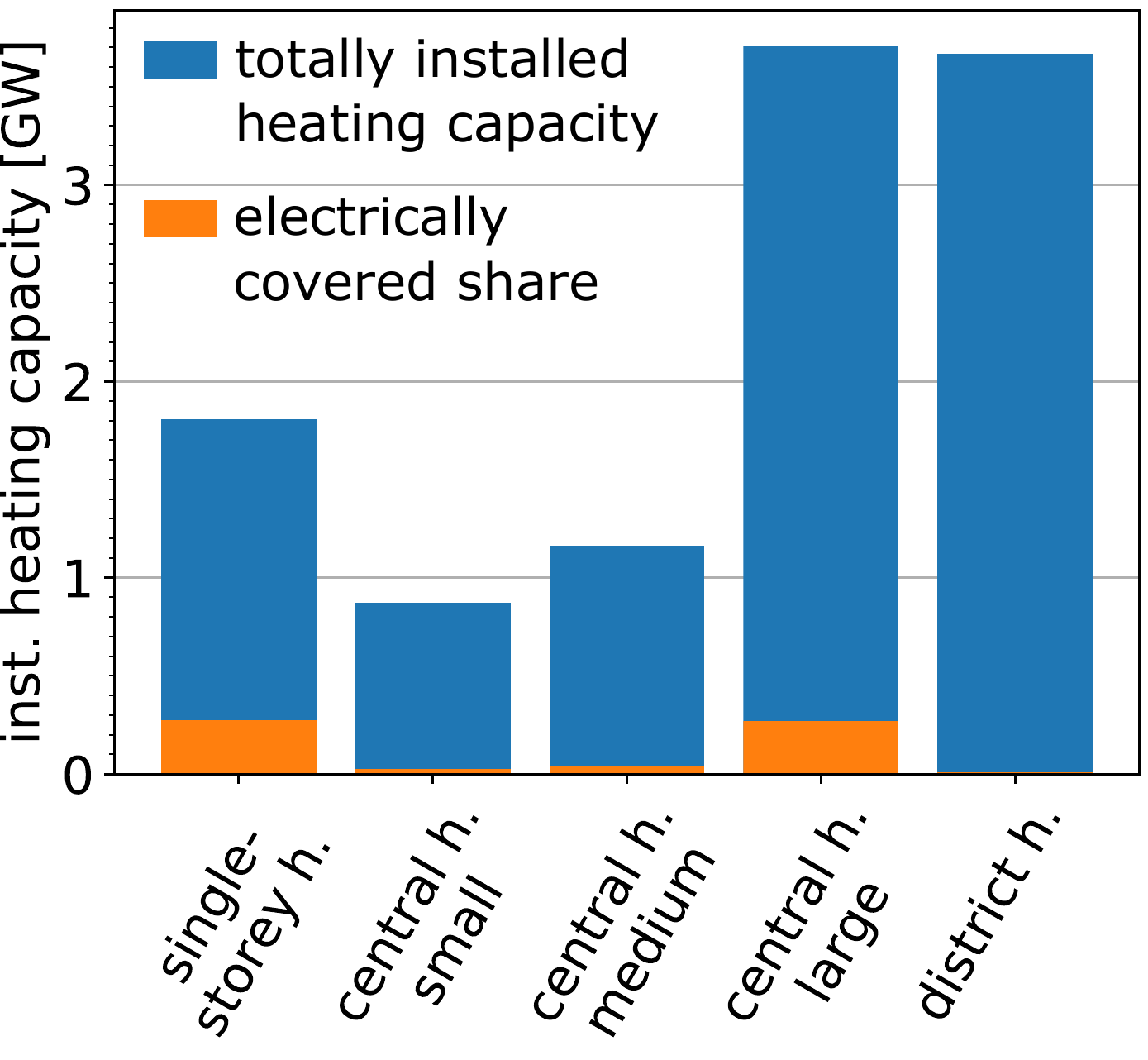}}
\hspace*{\fill}
\subfloat[\label{diag_berlin_hl_electric_share_tech_distribution}Shares of different technologies in the electrically covered share of the installed heating capacity.]{\includegraphics[width=0.475\textwidth]{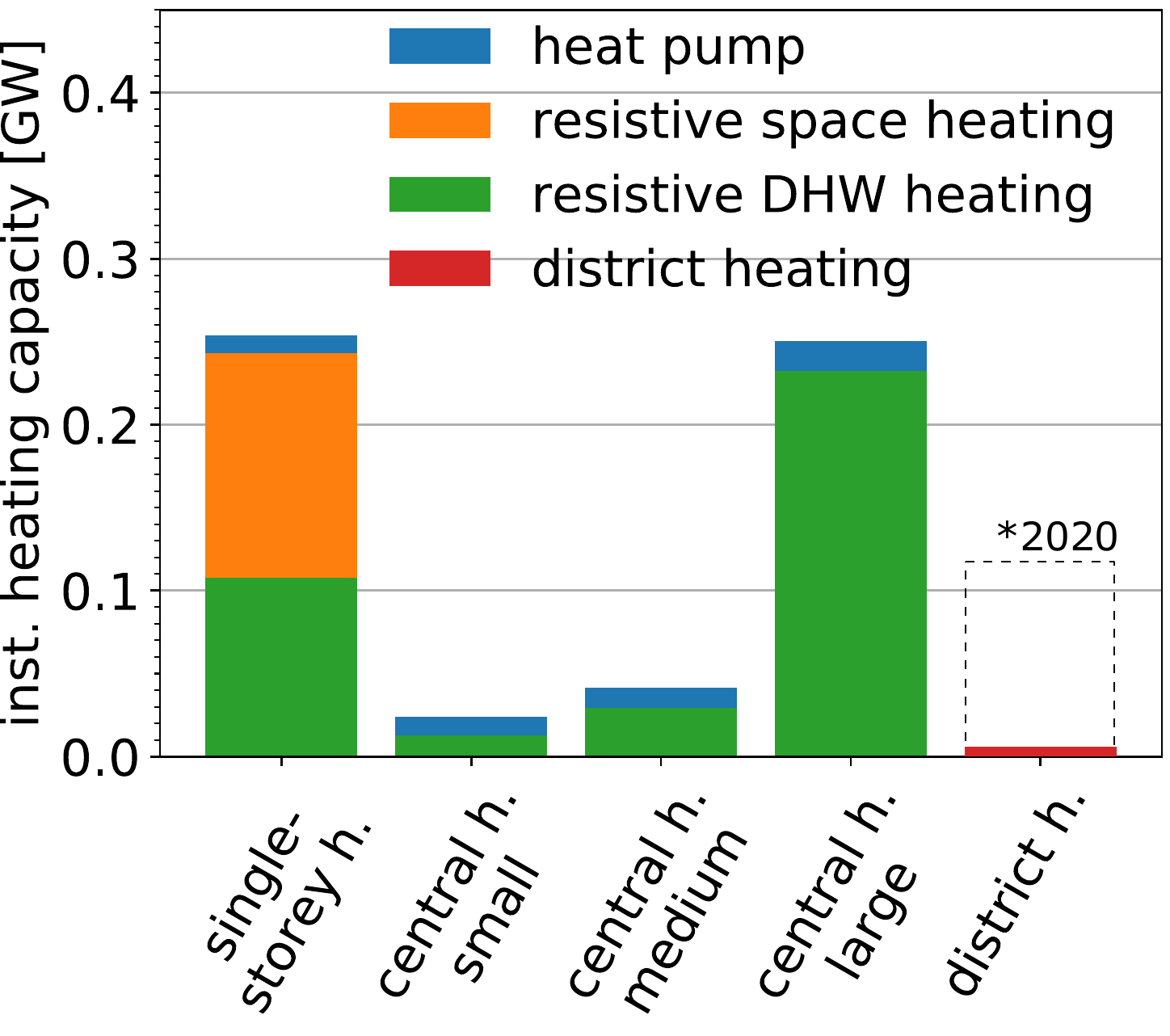}}
\caption{Installed heating capacity covered by electric heating technologies in the city of Berlin (subdivisions for central heating: small=$\dot{Q}_{inst}<12.5~kW_{th}$; medium=$12.5~kW_{th}<\dot{Q}_{inst}<25~kW_{th}$; large= $25~kW_{th}<\dot{Q}_{inst}$).}
\end{figure}

The results are depicted in Figure~\ref{diag_berlin_heat_load_electric_share} using the example of the city of Berlin.
The shares of the electrically covered load are relatively low, with a maximum of $0.25~GW_{th}$ from totally $1.8~GW_{th}$ heat load for the category single-storey heating, which corresponds to $13~\%$. 
Figure~\ref{diag_berlin_hl_electric_share_tech_distribution} shows, how the electrically covered load is split up into different technologies. The largest part is covered by resistive DHW heating, followed by resistive space heating. Heat pumps still play a minor role. Currently, there is one $6~MW_{th}$ resistive heater installed in the district heating system in Berlin. Compared to the cumulated electrically covered load by single-storey heatings it is a much lower value. The dashed black line represents a resistive heater under construction, as described in Section~\ref{sec_res_fut_scen}.

In Figure~\ref{fig_heating_class_size_distribution_GER}, the PtH facilities in district heating systems in Germany are presented. The smallest one, situated in the Spree-Neisse district, has a power of $0.55~MW_{th}$ and the largest one, situated in the city of Heilbronn, has a power of $100~MW_{th}$. According to the data provided in~\cite{christidis_2017_eneff}, only resistive heaters, no large scale heat pumps (like e.g. in Denmark~\cite{bach2016integration}) are installed in the district heating systems in Germany.

\subsection{Temporally resolved heat loads}
In this section, we present the results for the temporally resolved heat load. First, the intraday load profile for different weekdays is described. Then these profiles are superimposed with the ambient temperature dominated load profile in the course of the year. 

\subsubsection*{Intraday load profile}
\begin{figure}[h!]
\centering\includegraphics[width=1\linewidth]{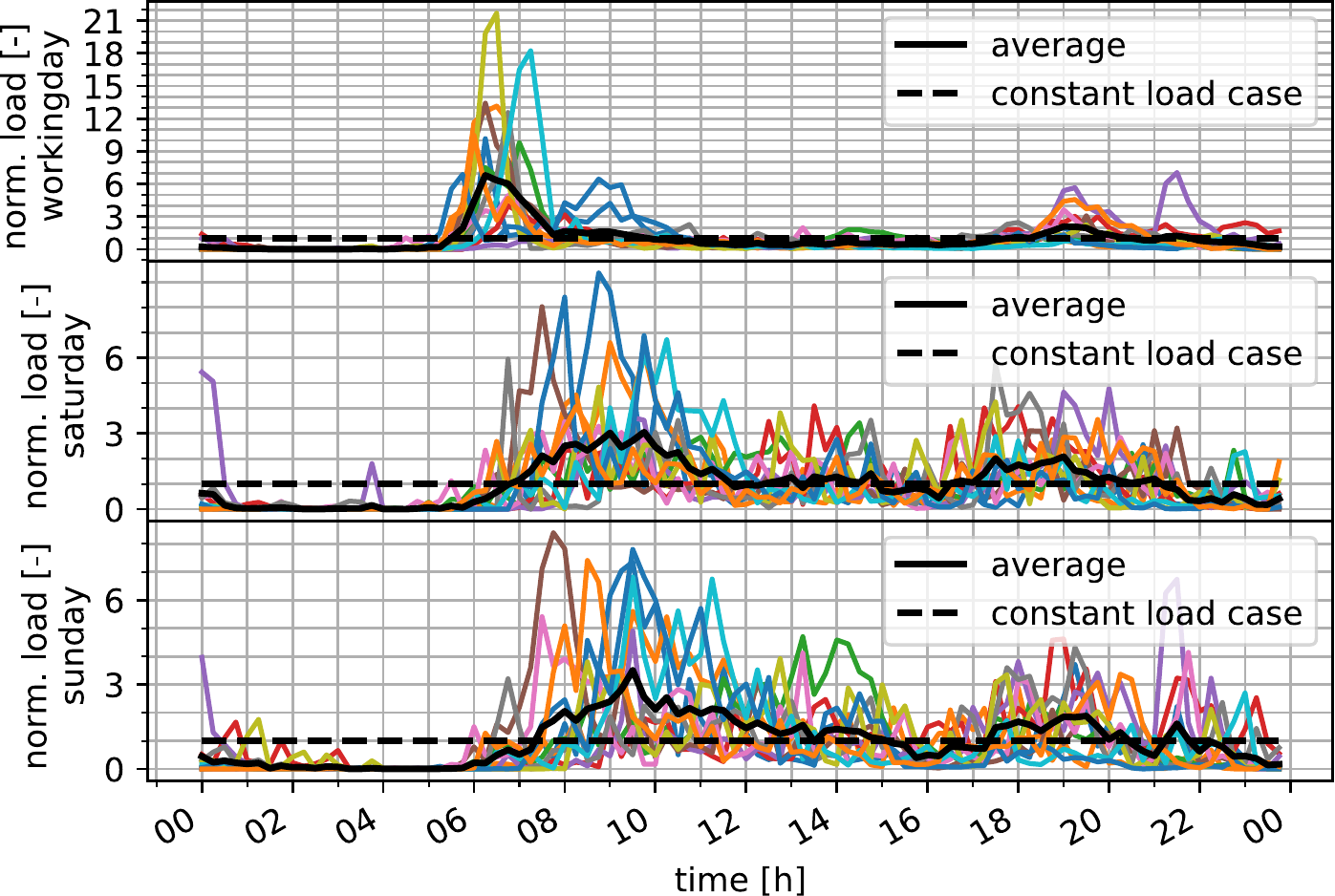}
\caption{Normalised DHW intraday load profile; for constant load case: $norm.~load~=~1$.}
\label{fig_intraday_profile_DHW}
\end{figure}

\begin{sloppypar} 
As described in Section~\ref{sec_meth_intraday_load_profile}, the intraday thermal load profiles have been derived from measurements that were carried out in the NOVAREF project~\cite{lange_2018_novaref}.
The results for DHW are shown in Figure~\ref{fig_intraday_profile_DHW}, where one coloured line corresponds to the average daily load profile of one building. The solid black line represents the average of the daily load profiles of all 12 buildings for which measurements were carried out. The dashed black line shows the reference case of a constant load, where the DHW demand is equally distributed over the entire day ($|{<}\dot{Q}_{15min}{>}|=1$). 
\end{sloppypar}

\begin{figure}[h!]
\centering\includegraphics[width=1\linewidth]{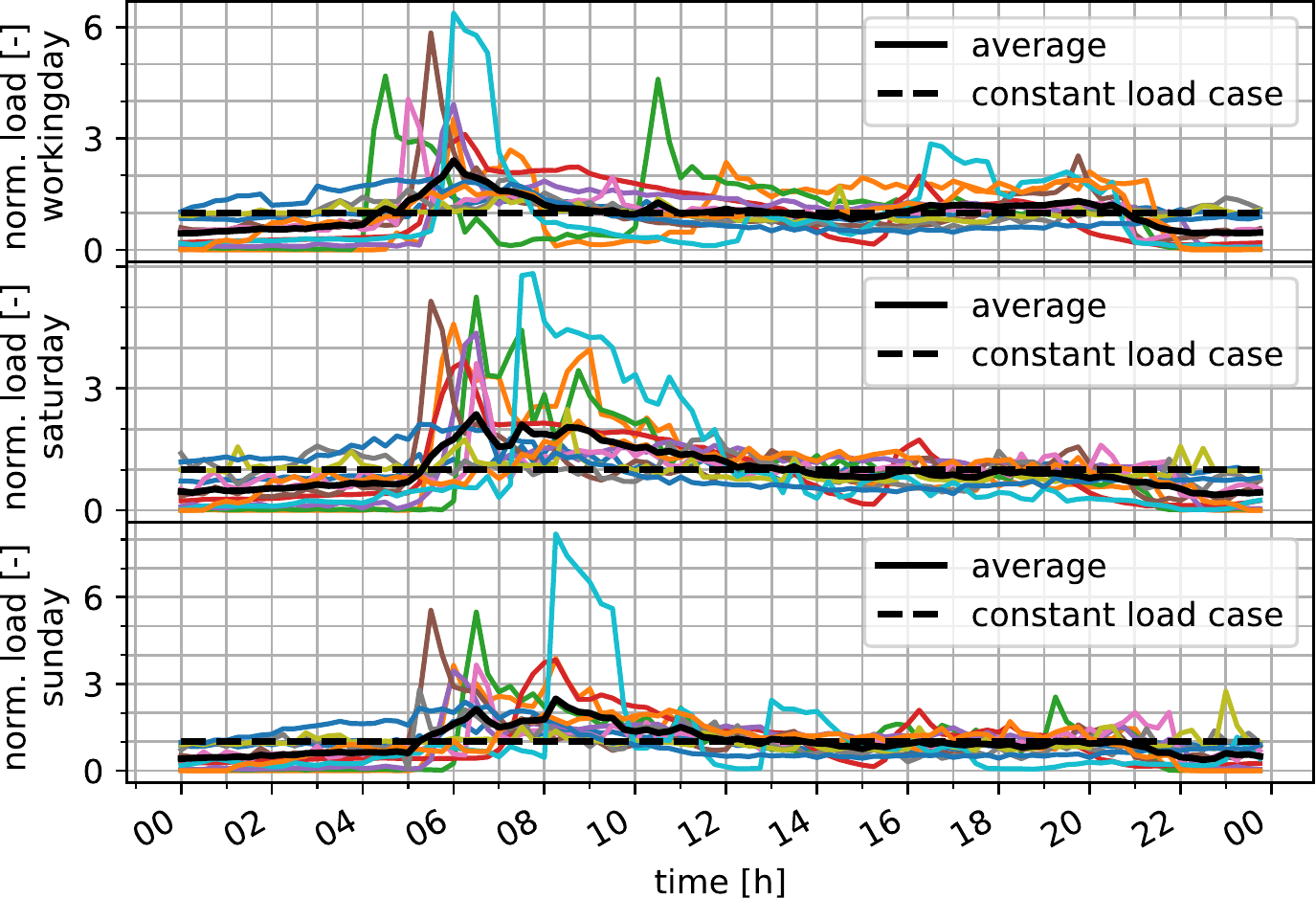}
\caption{Normalised space heating intraday load profile; for constant load case: $norm.~load~=~1$.}
\label{fig_intraday_profile_space_heat}
\end{figure}

On working days between 06:00 and 07:00, the average thermal load is six times higher than the constant load case. This is due to the use of DHW for showering and preparing breakfast with a high level of simultaneity.
There is a second, flatter load peak between 18:30 and 20:00, where the average load value is about two times higher than in the constant load case. The reason for this is probably, that most of the people come home after work in the evening hours and use DHW, but the level of simultaneity is lower than in the morning hours.

The DHW average load profile for Saturdays and Sundays also shows a morning and evening peak, which are flatter and span over a longer time than the one for working days. This  is due to the fact that weekends are non-working days and the load rather depends on the individual behaviour, leading to a lower level of simultaneity.

The load profile for space heating is shown in Figure~\ref{fig_intraday_profile_space_heat}. There is also a morning peak in the average load profile from 05:00 to 08:00 on working days, as well as from 05:30 till 12:00 on Saturdays and Sundays. The load at these peaks is approximately twice as high as for the constant load case. However, no significant evening peak is recognisable. Note also that, between 22:00 and 05:30 the load is about half of the constant load case for all of the three load profiles. The coloured profiles for the individual building show significant ramps, e.g. in the morning. This leads to the assumption that, the space heating profiles are only indirectly coupled to the demand of the residents. The profiles tend to be mainly dominated by the controller settings, e.g. starting the space heating at a specific point of time every morning. The numerical values of the average daily load time series for space heating and DHW are given in the supplementary material of this contribution\footnote{\label{note_supp_mat_results}\href{https://doi.org/10.5281/zenodo.2650200}{Supplementary material/results}}.

\subsubsection*{Yearly load profile}
\begin{figure}[h!]
\centering\includegraphics[width=\linewidth]{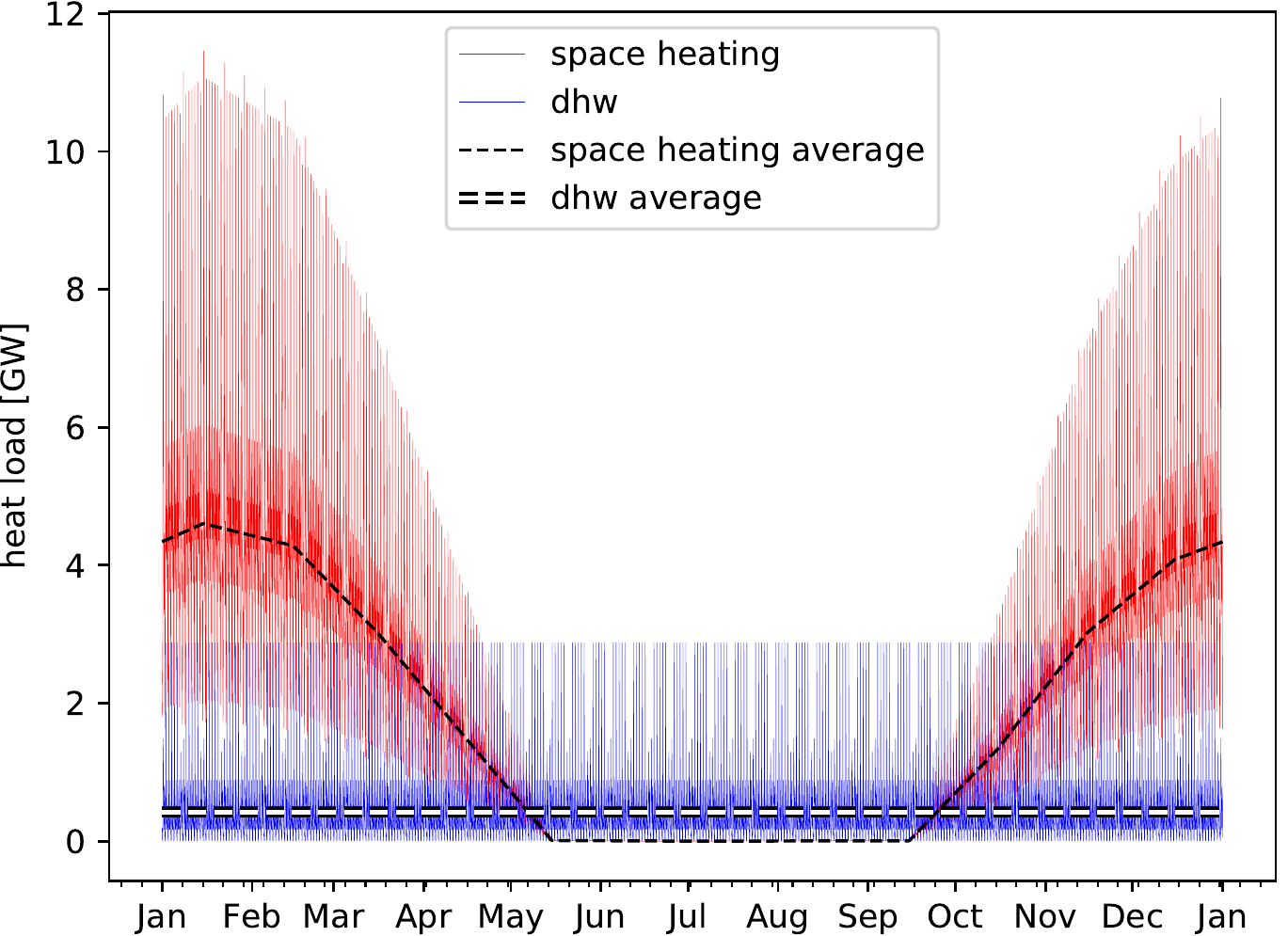}
\caption{Space heating and DHW load in the course of the year, calculated for the city of Berlin.}
\label{diag_heat_load_total_year_berlin_preliminary}
\end{figure}

The space heating and DHW load in the course of the whole year with a 15-minute resolution is shown in Figure~\ref{diag_heat_load_total_year_berlin_preliminary} for the example of Berlin.
In January, the space heating load has its maximum and then its value drops, till the ambient temperature reaches the heating limit temperature in May. During summer, only energy for DHW is needed. After the ambient temperature drops again below the heating limit temperature in September, the space heating load rises till January. During spring and autumn, the DHW load peaks are in the same order of magnitude as the average space heating load. The numerical values for the yearly load profiles of all administrative districts in Germany can be found in the supplementary material of this contribution\textsuperscript{\ref{note_supp_mat_results}}.

\subsection{Future scenario}\label{sec_res_fut_scen}
\begin{figure}[h!]
\subfloat[\label{berlin_installed_heat_load_future_projection}Totally installed heating capacity and the electrically covered share.]{\includegraphics[width=0.475\textwidth]{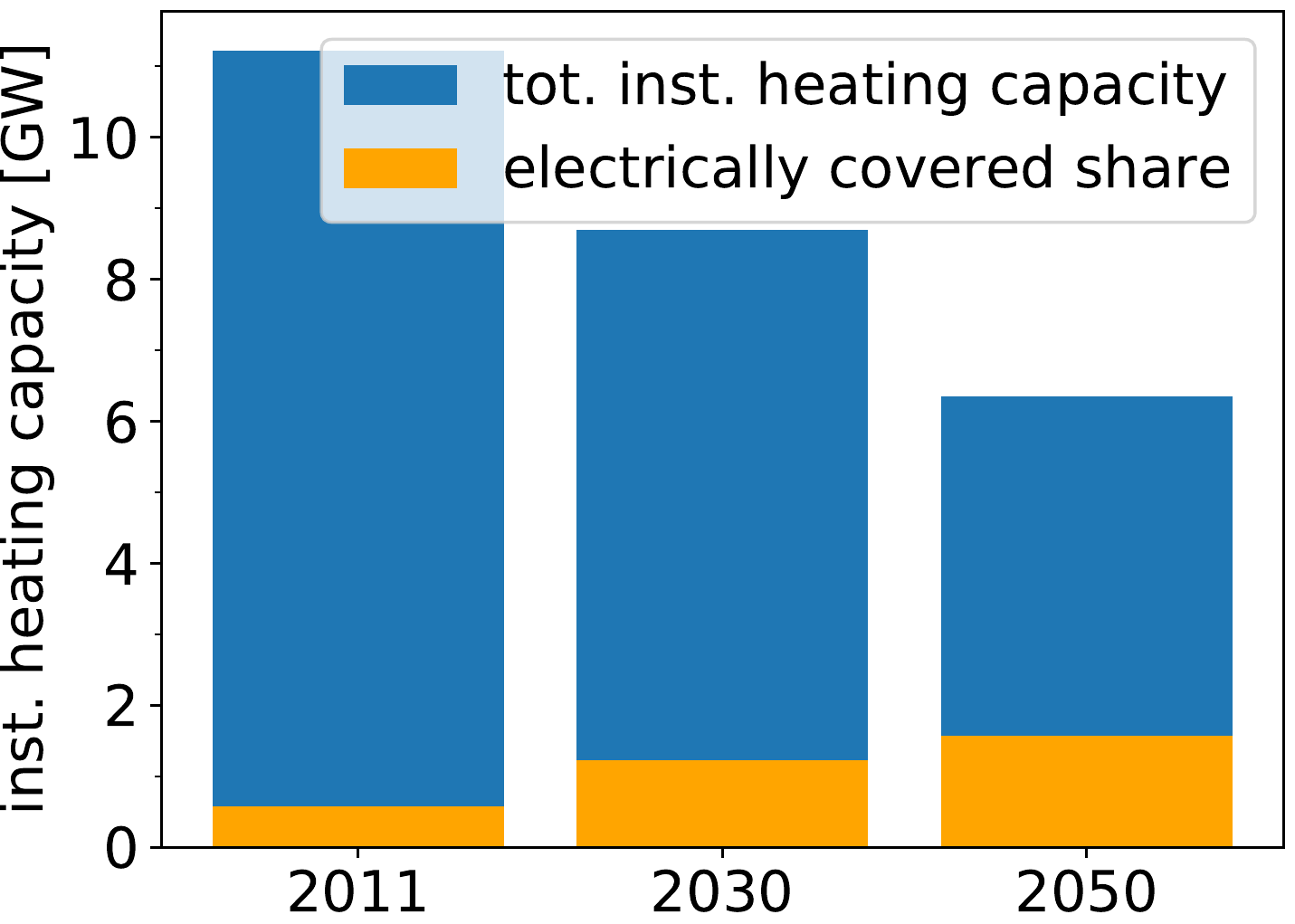}}
\hspace*{\fill}
\subfloat[\label{berlin_PtH_heat_dem_categories_tech_shares_future_projection}Technology shares in the electrically covered heat load.]{\includegraphics[width=0.475\textwidth]{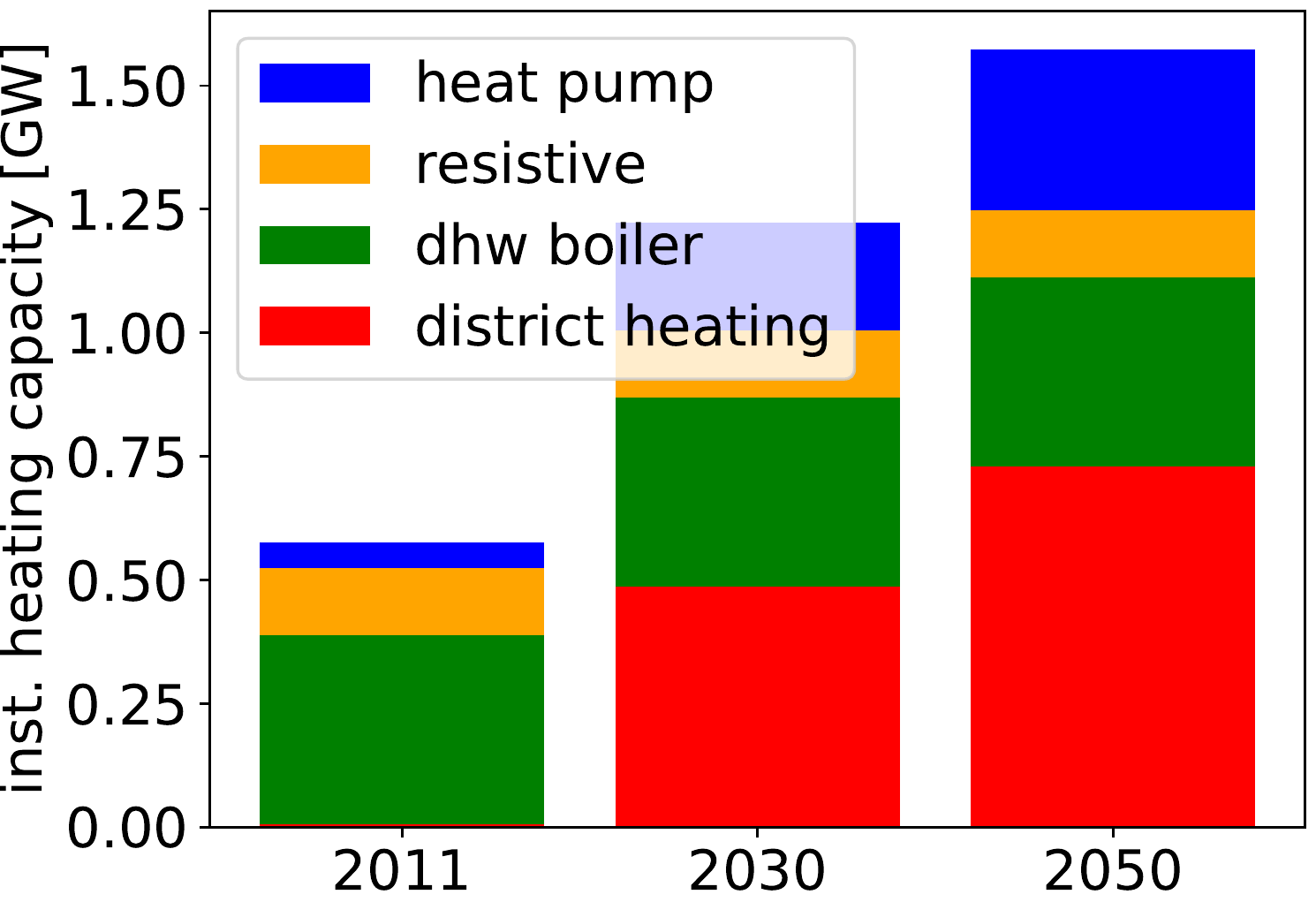}}
\caption{Future scenario for the installed heating capacity in Berlin.}
\end{figure}

The future development of the totally installed heating capacity in Berlin and the electrically covered share based on the values considered in Section~\ref{sec_meth_fut_dev}, is shown in Figure~\ref{berlin_installed_heat_load_future_projection}.
Due to the expected increase of energy efficiency, the total installed thermal load is predicted to drop by $45~\%$ till 2050. At the same time, the share of the electrically covered load increases from $4.5~\%$ to $25~\%$. The future development of the technology shares in the electrically covered heat load is shown in Figure~\ref{berlin_PtH_heat_dem_categories_tech_shares_future_projection}. As described in Section~\ref{sec_meth_fut_dev}, we supposed the overall installed load of resistive space heating devices and DHW boilers stays constant. For both 2030 and 2050, the expected load in Berlin from centralised PtH in district heating is twice as high as the load from decentralised heat pumps in residential buildings. 

As denoted by the dashed line in Figure~\ref{diag_berlin_hl_electric_share_tech_distribution}, a $120~MW_{th}$ resistive heater shall be installed in the Berlin district heating system in 2020~\cite{plazzo_2017_P2H_berlin,kiefert_2018_power}. It will then be the largest PtH facility in Germany. The electrically covered load in the district heating system will be in the same order of magnitude as the cumulated load of decentralised resistive space heatings in Berlin. This shows, how one large scale investment can significantly change the technology shares in the electrically covered load.

\subsection{Validation}
In this section, we compare the results of the present study with those found in the literature. First, the cumulated heat demand of all German administrative districts is addressed. Then we compare different regionalisation methods and finally deal with the installed PtH capacity of different technologies. 

\subsection*{Yearly heat demand for total Germany}
\begin{figure}[h!]
\centering\includegraphics[width=0.85\linewidth]{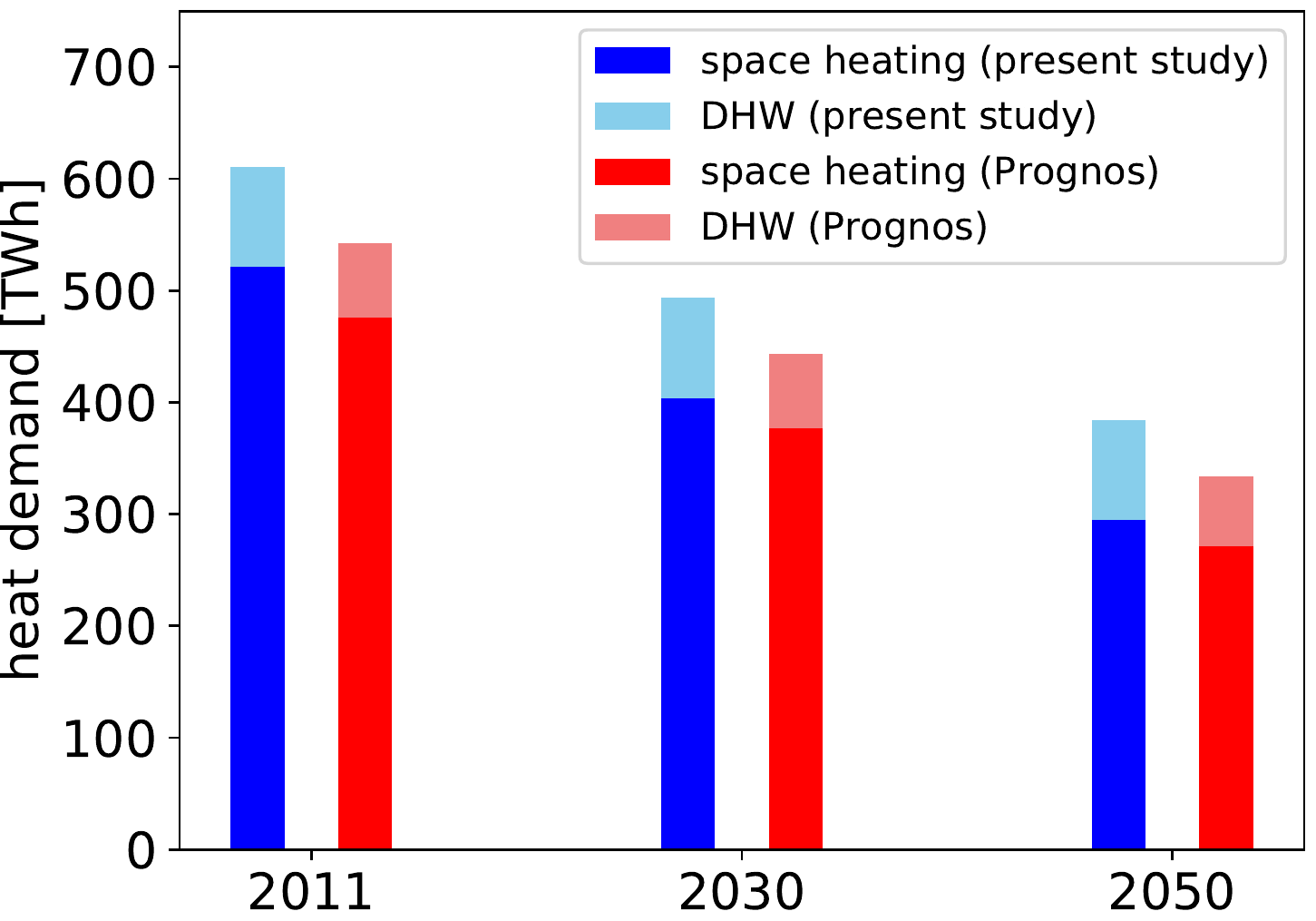}
\caption{Validation of the results for the yearly heat demand of total Germany with the data in~\cite{prognos_entwicklung_2014}.}
\label{fig_val_heat_dem_total_germany}
\end{figure}

We validated the results for the yearly space heating and DHW final energy demand for total Germany of the present study with the data in~\cite{prognos_entwicklung_2014}, as depicted in Figure~\ref{fig_val_heat_dem_total_germany}.
The DHW demand stays roughly constant from 2011 till 2050, with approximately $90~TWh$ in the present study and $70~TWh$ in~\cite{prognos_entwicklung_2014}. The discrepancy between these two values can be explained by different assumptions for the DHW usage of the German population. A comparison of DHW usage values is conducted in~\cite{ffe_2012_flex}.

The status quo space heating demand for Germany amounts to $520~TWh$ in this contribution and to $475~TWh$ in~\cite{prognos_entwicklung_2014}. This difference of $9~\%$ arises probably, because we used long term average ambient temperatures to calculate the heat demand and in~\cite{prognos_entwicklung_2014} data for 2011 were used only. In both studies, the yearly demand for space heating decreases by approx. $40~\%$ from 2011 till 2050.

\subsection*{Monofactorial and multifactorial regionalisation}
\begin{figure}[htbp]
\centering\includegraphics[width=0.85\linewidth]{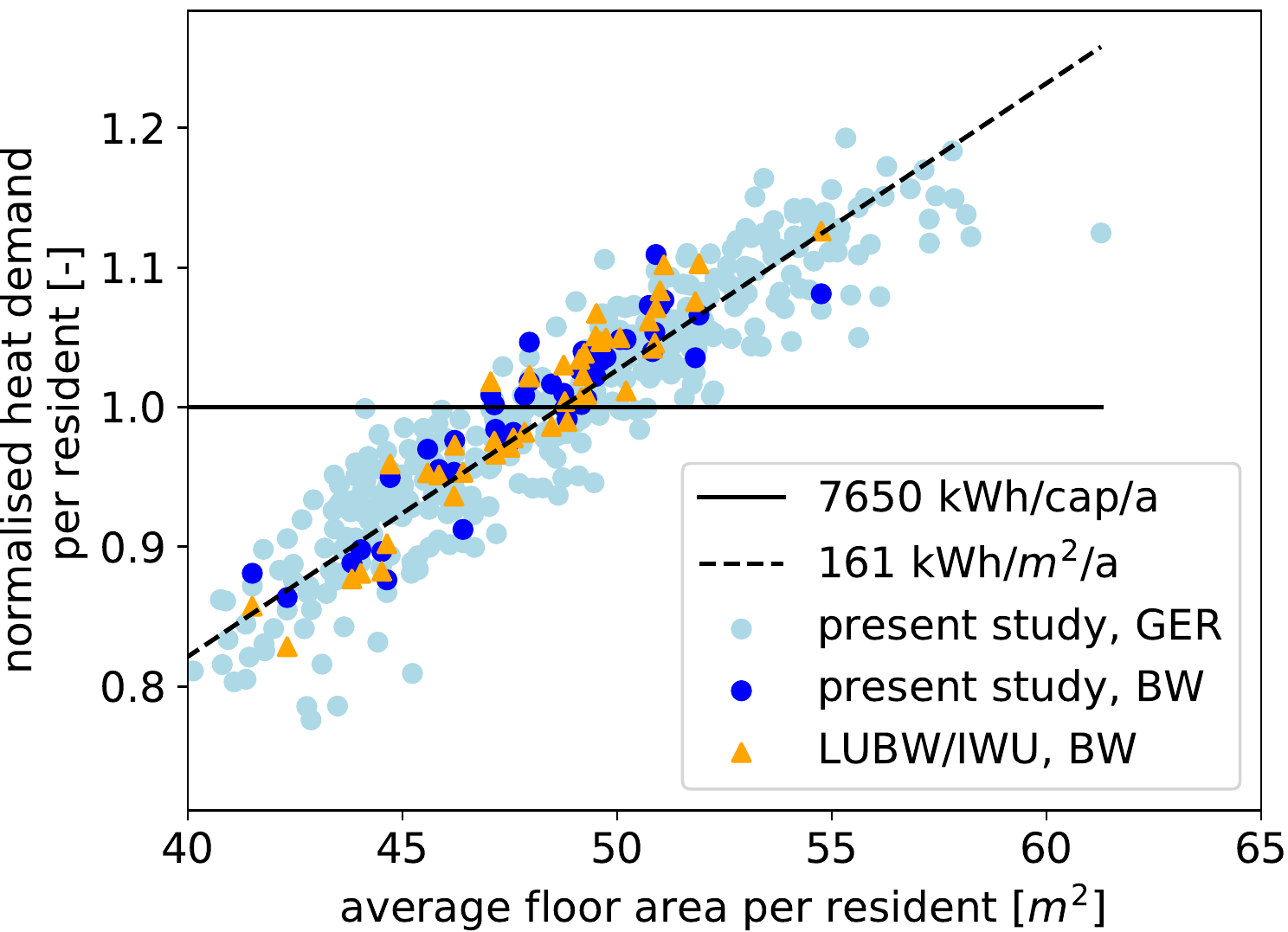}
\caption{Normalised heat demand per resident for all administrative districts of Germany~(GER) and the federal state Baden-Wuerttemberg~(BW).}
\label{diag_multi_fac_vs_mono_fac_regionalisation}
\end{figure}

In this paragraph, we compare the results of the present study with other methods to regionalise heat demand. We use Germany, respectively the state of Baden-Wuerttemberg as an example, and split up the total heat demand into single demand values for the associated administrative districts.
For a better comparability of the methods, the results of the regionalisation were normalised. As shown in Eq.~\ref{eq_q_ad_dist_norm}, we first divided the annual heat demand value $Q_{i}$ of each administrative district $i$ by the number of residents per district $n_{res,i}$. Second, we divided the resulting value $q_{res,i}$ by the average of the heat demand per resident of all administrative districts ${<}q_{res}{>}$:
\begin{equation}
|q_{res,i}|=\frac{Q_{i}}{n_{res,i}\cdot {<}q_{res}{>}}=\frac{q_{res,i}}{{<}q_{res}{>}}. \label{eq_q_ad_dist_norm} 
\end{equation}
In Figure~\ref{diag_multi_fac_vs_mono_fac_regionalisation}, the normalised annual heat demand per resident of the administrative districts $|q_{res,i}|$ is plotted over the average floor area per resident. One dot represents one administrative district. Thus the dots on the left side of the figure rather represent districts containing large cities, where a high share of the population lives in comparably small flats.
The dots on the right side rather represent rural areas, where many people live in rather large single-family houses.

A statistical parameter for which data are available in most cases, even for regions with a small extent~\cite{destatis2011census}, is the number of residents. The regionalisation using only the number of residents per region, is denoted by the solid black line in Figure~\ref{diag_multi_fac_vs_mono_fac_regionalisation}. We henceforth denote this kind of approach "monofactorial regionalisation" as only one statistical parameter is used.
In this method, the annual heat demand of Germany\footnote{\label{note_total_heat_dem_ger}We used the result from the present study, $610~TWh$.} is divided by the number of inhabitants\footnote{We used $80.21\cdot 10^6~inhabitants$, according to \href{https://www.zensus2011.de/DE/Home/home_node.html}{www.zensus2011.de}.}, yielding $7650~kWh/cap/a$. 
This constant is then multiplied with the number of inhabitants of each district, to get the heat demand per district. The subsequent normalisation yields a heat demand per resident of $|q_{res,i}|=1$ for every district.

However, literature agrees that the influence of the floor area on the heat demand is significantly higher than the one of the number of inhabitants~\cite{din2011_18599,din2008_12831,vdi2008_3807,bigalke2015dena_geb_rep,wei2014driving}.
So if statistical data on the floor area per district are available, another monofactorial regionalisation can be conducted, as indicated by the dashed line in Figure~\ref{diag_multi_fac_vs_mono_fac_regionalisation}. The annual heat demand of Germany\textsuperscript{\ref{note_total_heat_dem_ger}} is divided by the total floor area\footnote{We used $3.78\cdot 10^9~m^2$, according to \href{https://www.zensus2011.de/DE/Home/home_node.html}{https://www.zensus2011.de}.\\} in Germany, which yields $161~kWh/m^2/a$. This factor is then multiplied with the total floor area per district. Figure~\ref{diag_multi_fac_vs_mono_fac_regionalisation} shows that for the district with the lowest floor area per resident in Germany, the floor area specific regionalisation leads to an approx. $20\%$ lower heat energy demand than the resident specific regionalisation. For the district with the highest floor area per resident, the floor area specific regionalisation leads to an approx. $25\%$ higher energy demand than the resident specific regionalisation. Due to this high deviations, we do not recommend using the resident specific regionalisation. It may only be used in cases, when no other statistical data are available than the number of residents per district.

In the present study, more factors influencing the heat demand additionally to the floor area and number of residents were considered: the building type, number of flats per building, year of construction and the heating type. We thus denote this approach by "multifactorial regionalisation".
As shown in Figure~\ref{diag_multi_fac_vs_mono_fac_regionalisation}, also in the present study the normalised heat demand per resident of the administrative districts generally increases over the average floor area per resident. This indicates that the floor area is the main driver of the heat demand. Due to the additionally regarded building properties, the results of the present study differ from the ones of the floor area specific regionalisation, with maximal $+9~\%$  and minimal $-13~\%$.

The Baden-Wuerttemberg State Institute for the Environment (LUBW) regionalised the heat demand of the German State Baden-Wuerttemberg to its associated administrative districts\footnote{\href{http://udo.lubw.baden-wuerttemberg.de/projekte/pages/selector/index.xhtml;jsessionid=A968F0874077E7906BD5B3D9386551A3.projekte2}{For more information visit: https://www.lubw.baden-wuerttemberg.de}}~\cite{lubw_2016_energieatlas}. Also in~\cite{lubw_2016_energieatlas}, multiple factors were taken into account that influence the heat demand: the floor area, year of construction and building type. Figure~\ref{diag_multi_fac_vs_mono_fac_regionalisation} shows that the results of~\cite{lubw_2016_energieatlas} are generally corresponding to the results of the present study for Baden-Wuerttemberg and the Pearson Correlation Coefficient~\cite{lee1988thirteen} is $R=0.95$. 

Furthermore, we subtracted the results of the monofactorial floor area specific regionalisation approach from the results of both multifactorial regionalisation approaches, the present study and~\cite{lubw_2016_energieatlas}, taking into account all administrative districts of Baden-Wuerttemberg. The two resulting sets of deviations are plotted one versus the other in Figure~\ref{fig_val_heat_dem_deviation}~(see~\ref{app_dev_mono_mult}). The pearson correlation coefficient of the deviations is $R=0.55$, which stands for a moderate correlation. That means that the results of the multifactorial regionalisation methods (the present study and~\cite{lubw_2016_energieatlas}) do not just randomly, but systematically differ from the monofactorial regionalisation method.

\subsection*{Installed PtH capacity of different technologies}
To validate the installed heating capacity of different PtH technologies, we compared the results of the present study with those obtained in~\cite{ffe_merit_2016}.
As the data in~\cite{ffe_merit_2016} were given as installed electrical capacity, we converted them to thermal capacities. The electric capacity of heat pumps was multiplied with an annual coefficient of performance of 2.89, that was also given in~\cite{ffe_merit_2016}. Regarding the resistive heating technologies used for space heating, DHW heating and in district heating systems, we approximated that the thermal capacity is equal to the electrical capacity. Furthermore in~\cite{ffe_merit_2016} it is distinguished between the amount of PtH capacity that can be additionally activated at a certain point of time and the one, which can be deactivated. We selected the respective maximum of these values, to compare it with the installed heat load, determined in the present study.

\begin{figure}[h!]
\centering\includegraphics[width=0.85\linewidth]{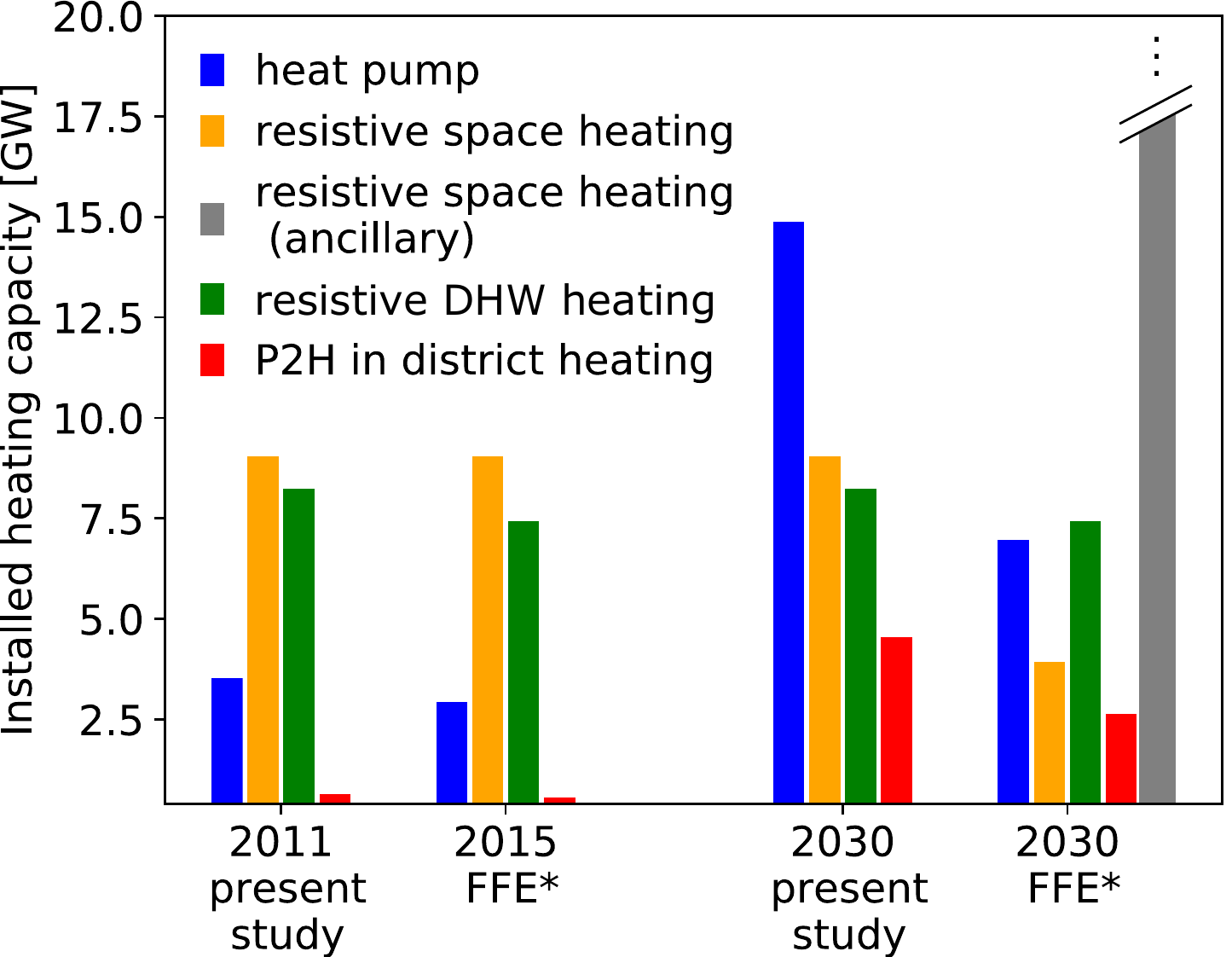}
\caption{Validation of the results for the PtH potentials for total Germany, *data from~\cite{ffe_merit_2016} converted as described in the text.}
\label{fig_val_PtH_total_germany}
\end{figure}
As shown in Figure~\ref{fig_val_PtH_total_germany}, the installed thermal capacity of heat pumps for the status quo is about the same size, with $3.5~GW$ in the present study and $2.89~GW$ in~\cite{ffe_merit_2016}. The thermal heat pump capacity will increase to $15~GW$ by 2030 in the present study and to $7~GW$ in~\cite{ffe_merit_2016}. This discrepancy can probably be explained by the different scenarios used for the future heat pump expansion.

The installed capacity of resistive space heatings amounts to $9~GW$ for the status quo in both studies. While the capacity stays constant in the present study till 2030, it drops to $4~GW$ in~\cite{ffe_merit_2016}. This difference arises because the authors of~\cite{ffe_merit_2016} did not take into account, that the need for balancing of renewable energies may have an increasing effect on the numbers of installed space heatings. Instead they considered an additional category of ancillary resistive heatings that may be combined with a conventional heating system, that runs on fossil fuels. The authors of~\cite{ffe_merit_2016} estimate the potential of this category at $100~GW$ by 2030. The installed resistive DHW heating capacity amounts to $8.2~GW$ in the present study and $7.4~GW$ in~\cite{ffe_merit_2016}. In both studies the capacity remains constant till 2030. For the status quo, also the capacity of resistive heating in district heating is approximately equal in both studies, with $0.6~GW$. Regarding the 2030 horizon, the PtH capacity will increase to $4.5~GW$ in the present contribution and to $2.6~GW$ in~\cite{ffe_merit_2016}. As the future expansion of this PtH technology strongly depends on single large scale investments of district heating operators, it can be hardly predicted.

\section{Conclusion and Outlook}\label{sec_conclusion_outlook}

With the expansion of renewable energies in Germany, events of imminent grid congestion occur more often. In order to avoid the curtailment of renewable energy sources, one option is to use excess feed-in locally for heating applications.    
As a first step to assess the potential of converting excess power to heat, in this contribution we determined the overall heat demand and the electrically covered share in the residential building sector. To account for regional differences, the potentials were spatially resolved on the administrative district level.    

In contrast to the other studies in this field, we provide all results as open data and take a higher number of building attributes into account for the regionalisation of the building stock. For this purpose, a special evaluation of the census enumeration data was ordered at the Research Data Centre of the German Federal Statistical Office, with a cross combination of six building attributes that influence the heat demand. Using these data, 729 building categories were generated and the number of buildings per category for each administrative district was determined. For each building category, heat demand values, as well as daily and yearly load profiles were assigned that were generated from measurement values.

We distinguished between different heating technologies and three classes of installed heating capacity per building. For urban districts, this yielded a high installed heating capacity of small scale single-storey heatings, as well as large central heatings. Both is due to the high share of multi-family houses in cities. In rural areas there are more small scale and medium scale central heating systems, because of the high number of single family houses.  Most district heating networks were found in large cities but also a not negligible capacity in rather rural administrative districts.

For the electrically covered heat load, we took into account decentralised heat pumps, resistive space heating and resistive DHW heating. Also centralised resistive heating facilities in district heating systems were considered. Thus the distribution of the installed capacity on the different PtH technologies in the administrative districts could be demonstrated. At the example of Berlin it was shown, how one large scale investment can significantly change the shares of the different technologies.

A generalized future scenario was defined, taking into account the 2030 and 2050 horizon. In this scenario the overall heat demand of Germany decreases by $40~\%$ from 2011 till 2050, while the electrically covered share increases by a factor of five.
Moreover, different approaches to regionalise the overall heat demand of Germany to its respective administrative districts were compared. Taking into account only one factor for the regionalisation, as e.g. the population density, results in differences up to $25~\%$ from the present study. We therefore recommend to consider multiple building attributes for the regionalisation of the building stock and heat demand. 

The regionalised heat demand and power-to-heat potentials determined in this contribution are valuable input data for future research. As the results are provided as open data, they can be used by other researchers for assessing the contribution of the power-to-heat technology to RES integration. When investigating the flexible operation of PtH devices, the heating capacity classes that were introduced in this contribution are beneficial. The specific costs for heat storage and for equipping the PtH devices with ICT generally drop with larger installed heating capacity, which may influence the economic viability.

The future scenario that was regarded in this contribution may be further extended. The high number of defined building categories, allows for applying individual development forecasts for the different building types. Also more details on new construction and demolition rates of residential buildings and reinvestment cycles of heating systems may be included.

\section*{Acknowledgements}
The first author gratefully acknowledges the financial support provided by the Foundation of German Business (sdw) through a PhD scholarship. The authors also thank Marco Zobel, Michael Lange, Peter Klement and Christoph Schillings for the fruitful discussions on residential heat demand modelling.

\appendix
\section{Supplementary Material}\label{app_supp_mat}
The residential heat demand and power-to-heat potential data for all administrative districts in Germany (NUTS-3) determined in this contribution, the used input data, as well as the developed source code are provided in the supplementary material of this contribution under open source licenses. The Data can be found on the zenodo repository at: \href{https://doi.org/10.5281/zenodo.2650200}{https://doi.org/10.5281/zenodo.2650200}.

\section{Implementation of the heat demand and PtH potential determination}\label{app_implementation}

The determination of the regionalised heat demand and PtH potentials is implemented in PostgreSQL.
The code was executed under a Debian GNU / Linux 9 environment, using PostgreSQL 9.4.18.
\begin{figure}[h]
\centering\includegraphics[width=\linewidth]{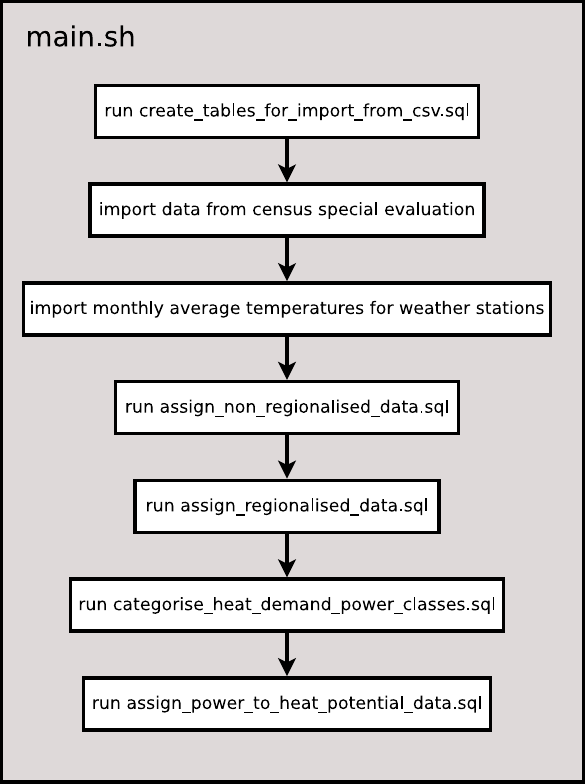}
\caption{The bash script main.sh calls the sub-processes for determining the regionalised heat demand and PtH potentials.}\label{fig_app_main_sh}
\end{figure}
As shown in Figure~\ref{fig_app_main_sh}, the process is divided into multiple SQL files that are called by the bash script \verb|main.sh|. 
First, the required SQL tables are created and input data are imported. In the file \verb|assign_non_regionalised_data.sql|, the heat demand values are assigned to the building categories. The heat demand is then regionalised to the administrative districts with the execution of the file \verb|assign_regionalised_data.sql|. Finally, different classes of installed heating capacity are introduced and the PtH potentials are assigned by running the files \verb|categorise_heat_demand_power_classes.sql| and \verb|assign_power_to_heat_potential_data.sql|. The developed source code and a more comprehensive flowchart of the sub-processes are provided in the supplementary material of this contribution\footnote{\href{https://doi.org/10.5281/zenodo.2650200}{Supplementary material/code}}.

\section{Influence of building type on heat demand}\label{app_influ_build_type}

By Eq.~\ref{eq:heat_dem_a_v_dep} it can be calculated, how the building type influences the heat demand of buildings. When disaggregating the demand values, obtained from~\cite{bigalke2015dena_geb_rep}, according to the building type, it must be ensured that the overall heat demand of all building type categories does not change. This is expressed by the following equation:     
\begin{align}\label{eq_conservation_disaggregation}
q''\cdot (A_{det}+A_{sd}+A_{row})=q''_{det}\cdot A_{det}+q''_{sd}\cdot A_{sd}+q''_{row}\cdot A_{row}, 
\end{align}
where q'' stands for the aggregated area specific heat demand value of all building types, $q''_{det}$ for the disaggregated area specific heat demand value of all detached buildings, $q''_{sd}$ for the demand of the semi-detached buildings and $q''_{row}$ for the demand of the row type buildings, $A_{det}$ stands for the cumulated floor area of all detached buildings, $A_{sd}$ for the area of the semi-detached buildings and $A_{row}$ for the area of the row buildings. The Eq.~\ref{eq_conservation_disaggregation} is then transformed to get the ratios of the disaggregated demand values to the aggregated values, as follows,
\begin{align}
\frac{q''_{det}}{q''}&=\frac{A_{det}+A_{sd}+A_{row}}{A_{det}+A_{sd}\cdot q''_{sd}/q''_{det}+A_{row}\cdot q''_{row}/q''_{det}},\label{eq_q_ratios_1}\\
\frac{q''_{sd}}{q''}&=\frac{A_{det}+A_{sd}+A_{row}}{A_{det}\cdot q''_{det}/q''_{sd}+A_{sd}+A_{row}\cdot q''_{row}/q''_{sd}},\label{eq_q_ratios_2}\\ 
\frac{q''_{row}}{q''}&=\frac{A_{det}+A_{sd}+A_{row}}{A_{det}\cdot q''_{det}/q''_{row}+A_{sd}\cdot q''_{sd}/q''_{row}+A_{row}}.\label{eq_q_ratios_3} 
\end{align}
The denominator of the equations~\ref{eq_q_ratios_1}, \ref{eq_q_ratios_2} and~\ref{eq_q_ratios_3} contains the ratios of the disaggregated heat demands of different building types. We calculated these ratios by applying Eq.~\ref{eq:heat_dem_a_v_dep} for the respective building types and dividing the results by each other. The results of the equations~\ref{eq_q_ratios_1}, \ref{eq_q_ratios_2} and~\ref{eq_q_ratios_3} were then each multiplied with the nine area specific heat demand values obtained from~\cite{bigalke2015dena_geb_rep} depending on the year of construction and number of flats per building, which yields 27 disaggregated values depending additionally on the building type. The numerical values for the process described here, can be found in the supplementary material of this contribution\footnote{\href{https://doi.org/10.5281/zenodo.2650200}{Supplementary material/other input data/heat demand according to building type year of construction and number of flats per building}}.

\section{Heat load curves for building categories}\label{app_heat_load_curves}
\begin{figure}[h]
\centering\includegraphics[width=0.85\linewidth]{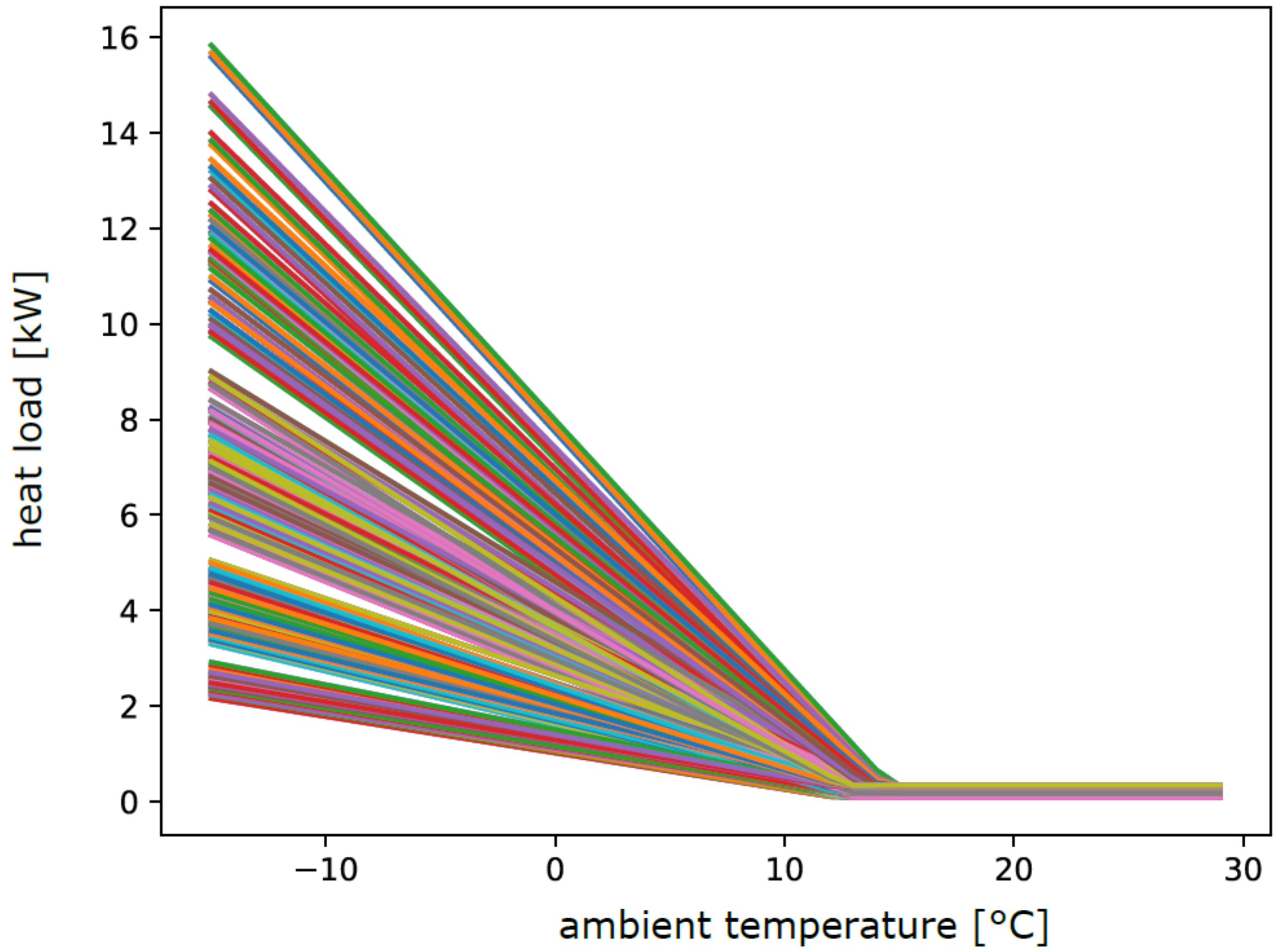}
\caption{Heat load as a function of the ambient temperature for all building categories defined in Section~\ref{sec_meth_spatial_mod} (one colour represents one building category).}\label{fig_Q_categories_full}
\end{figure}

\section{PtH technology scaling factor for German states}\label{app_PtH_state_scal_fac}
The scaling factor $x_{ij}$ is defined for each of the technologies $i=1...3$ (heat pump, resistive space heating and resistive DHW heating) and for all German states $j$, as:
\begin{align}
x_{ij}&=\frac{n_{devices,ij}/n_{buildings,j}}{n_{devices,Germany,i}/n_{buildings,Germany}},\\
n_{devices,Germany,i}&=\sum_{j=1...16} n_{devices,ij},\\
n_{buildings,Germany}&=\sum_{j=1...16} n_{buildings,j}.
\end{align}
Therein $n_{devices,ij}$ is the number of installed PtH devices per state (data from~\cite{bwp_waermepump_2016} for heat pumps and from~\cite{destatis_microcen_2010} for resistive space heating and resistive DHW heating), $n_{buildings,j}$ is the number of buildings per state~\cite{census_sonderauswertung}, $n_{devices,Germany,i}$ is the number of installed PtH devices for total Germany and $n_{buildings,Germany}$ is the number of buildings for total Germany. The numerical values of the scaling factors are provided in the supplementary material of this contribution\footnote{\label{foot_note_el_heat_fac_no2}\href{https://doi.org/10.5281/zenodo.2650200}{Supplementary material/other input data/electric heating factors}}.

\section{Heat storages in district heating systems}\label{app_heat_stor_dist_heat}
Heat storage is not in the focus of this contribution. However, the data on thermal storage capacities in district heating systems, listed in~\cite{christidis_2017_eneff}, may be very useful for researchers, investigating the time shifting of PtH operation. We thus also assigned these capacity data to the respective administrative districts and added them to Figure~\ref{fig_heating_class_size_distribution_GER}. The numerical values are provided in the supplementary material of this contribution\textsuperscript{\ref{foot_note_el_heat_fac_no2}}. For several heat storage facilities, no thermal storage capacity was given in~\cite{christidis_2017_eneff}. We therefore calculated an average volume specific capacity of all the storages, for which data are given. This yielded $0.04~MWh_{th}/m^3$ for both the atmospheric and the 2-zone storage type and $0.07~MWh_{th}/m^3$ for the pressurised heat storage. We multiplied these average values with the volume of the storages for which no thermal capacity data were given.


\section{Deviation of the monofactorial regionalisation from different multifactorial regionalisation methods}\label{app_dev_mono_mult}
\begin{figure}[h!]
\centering\includegraphics[width=0.85\linewidth]{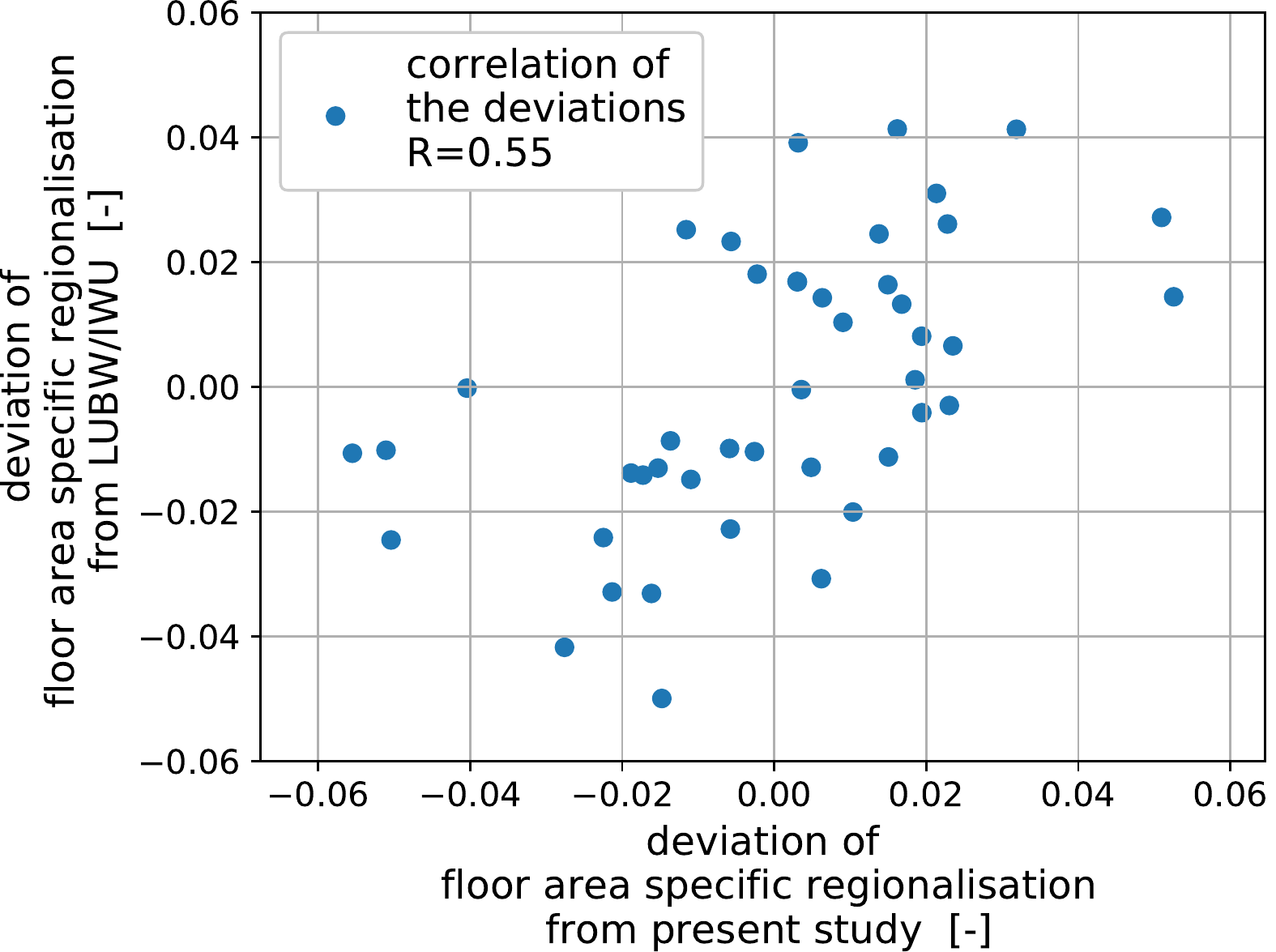}
\caption{Deviation of the normalised heat demand of the administrative districts calculated by the  monofactorial floor area specific regionalisation from the multifactorial regionalisation methods (x-axis: present study, y-axis: LUBW/IWU~\cite{lubw_2016_energieatlas})}
\label{fig_val_heat_dem_deviation}
\end{figure}




\bibliographystyle{model1-num-names}
\bibliography{main.bib}

\begin{thebibliography}{59}
\expandafter\ifx\csname natexlab\endcsname\relax\def\natexlab#1{#1}\fi
\providecommand{\bibinfo}[2]{#2}
\ifx\xfnm\relax \def\xfnm[#1]{\unskip,\space#1}\fi
\bibitem[{Sims(2004)}]{sims2004renewable}
\bibinfo{author}{R.~E. Sims},
\newblock \bibinfo{title}{{Renewable energy: a response to climate change}},
\newblock \bibinfo{journal}{Solar energy} \bibinfo{volume}{76}
  (\bibinfo{year}{2004}) \bibinfo{pages}{9--17}.
\bibitem[{Bloess et~al.(2018)Bloess, Schill, and Zerrahn}]{bloess2018power}
\bibinfo{author}{A.~Bloess}, \bibinfo{author}{W.-P. Schill},
  \bibinfo{author}{A.~Zerrahn},
\newblock \bibinfo{title}{{Power-to-heat for renewable energy integration: A
  review of technologies, modeling approaches, and flexibility potentials}},
\newblock \bibinfo{journal}{Applied Energy} \bibinfo{volume}{212}
  (\bibinfo{year}{2018}) \bibinfo{pages}{1611--1626}.
\bibitem[{Hake et~al.(2015)Hake, Fischer, Venghaus, and
  Weckenbrock}]{hake2015german}
\bibinfo{author}{J.-F. Hake}, \bibinfo{author}{W.~Fischer},
  \bibinfo{author}{S.~Venghaus}, \bibinfo{author}{C.~Weckenbrock},
\newblock \bibinfo{title}{{The German Energiewende--history and status quo}},
\newblock \bibinfo{journal}{Energy} \bibinfo{volume}{92} (\bibinfo{year}{2015})
  \bibinfo{pages}{532--546}.
\bibitem[{Bauermann(2016)}]{bauermann2016german}
\bibinfo{author}{K.~Bauermann},
\newblock \bibinfo{title}{{German Energiewende and the heating market--Impact
  and limits of policy}},
\newblock \bibinfo{journal}{Energy Policy} \bibinfo{volume}{94}
  (\bibinfo{year}{2016}) \bibinfo{pages}{235--246}.
\bibitem[{Hoogwijk et~al.(2005)Hoogwijk, Faaij, Eickhout, de~Vries, and
  Turkenburg}]{hoogwijk2005potential}
\bibinfo{author}{M.~Hoogwijk}, \bibinfo{author}{A.~Faaij},
  \bibinfo{author}{B.~Eickhout}, \bibinfo{author}{B.~de~Vries},
  \bibinfo{author}{W.~Turkenburg},
\newblock \bibinfo{title}{{Potential of biomass energy out to 2100, for four
  IPCC SRES land-use scenarios}},
\newblock \bibinfo{journal}{Biomass and Bioenergy} \bibinfo{volume}{29}
  (\bibinfo{year}{2005}) \bibinfo{pages}{225--257}.
\bibitem[{Schweiger et~al.(2017)Schweiger, Rantzer, Ericsson, and
  Lauenburg}]{schweiger2017potential}
\bibinfo{author}{G.~Schweiger}, \bibinfo{author}{J.~Rantzer},
  \bibinfo{author}{K.~Ericsson}, \bibinfo{author}{P.~Lauenburg},
\newblock \bibinfo{title}{{The potential of power-to-heat in Swedish district
  heating systems}},
\newblock \bibinfo{journal}{Energy} \bibinfo{volume}{137}
  (\bibinfo{year}{2017}) \bibinfo{pages}{661--669}.
\bibitem[{Averfalk et~al.(2017)Averfalk, Ingvarsson, Persson, Gong, and
  Werner}]{averfalk2017large}
\bibinfo{author}{H.~Averfalk}, \bibinfo{author}{P.~Ingvarsson},
  \bibinfo{author}{U.~Persson}, \bibinfo{author}{M.~Gong},
  \bibinfo{author}{S.~Werner},
\newblock \bibinfo{title}{{Large heat pumps in Swedish district heating
  systems}},
\newblock \bibinfo{journal}{Renewable and Sustainable Energy Reviews}
  \bibinfo{volume}{79} (\bibinfo{year}{2017}) \bibinfo{pages}{1275--1284}.
\bibitem[{B{\"o}ttger et~al.(2014)B{\"o}ttger, G{\"o}tz, Lehr, Kondziella, and
  Bruckner}]{bottger2014potential}
\bibinfo{author}{D.~B{\"o}ttger}, \bibinfo{author}{M.~G{\"o}tz},
  \bibinfo{author}{N.~Lehr}, \bibinfo{author}{H.~Kondziella},
  \bibinfo{author}{T.~Bruckner},
\newblock \bibinfo{title}{{Potential of the power-to-heat technology in
  district heating grids in Germany}},
\newblock \bibinfo{journal}{Energy Procedia} \bibinfo{volume}{46}
  (\bibinfo{year}{2014}) \bibinfo{pages}{246--253}.
\bibitem[{B{\"o}ttger et~al.(2015)B{\"o}ttger, G{\"o}tz, Theofilidi, and
  Bruckner}]{bottger2015control}
\bibinfo{author}{D.~B{\"o}ttger}, \bibinfo{author}{M.~G{\"o}tz},
  \bibinfo{author}{M.~Theofilidi}, \bibinfo{author}{T.~Bruckner},
\newblock \bibinfo{title}{{Control power provision with power-to-heat plants in
  systems with high shares of renewable energy sources--An illustrative
  analysis for Germany based on the use of electric boilers in district heating
  grids}},
\newblock \bibinfo{journal}{Energy} \bibinfo{volume}{82} (\bibinfo{year}{2015})
  \bibinfo{pages}{157--167}.
\bibitem[{Nyborg and R{\o}pke(2015)}]{nyborg2015heat}
\bibinfo{author}{S.~Nyborg}, \bibinfo{author}{I.~R{\o}pke},
\newblock \bibinfo{title}{{Heat pumps in Denmark: From ugly duckling to white
  swan}},
\newblock \bibinfo{journal}{Energy Research \& Social Science}
  \bibinfo{volume}{9} (\bibinfo{year}{2015}) \bibinfo{pages}{166--177}.
\bibitem[{Zhang et~al.(2016)Zhang, Lu, McElroy, Nielsen, Chen, Deng, and
  Kang}]{zhang2016reducing}
\bibinfo{author}{N.~Zhang}, \bibinfo{author}{X.~Lu}, \bibinfo{author}{M.~B.
  McElroy}, \bibinfo{author}{C.~P. Nielsen}, \bibinfo{author}{X.~Chen},
  \bibinfo{author}{Y.~Deng}, \bibinfo{author}{C.~Kang},
\newblock \bibinfo{title}{{Reducing curtailment of wind electricity in China by
  employing electric boilers for heat and pumped hydro for energy storage}},
\newblock \bibinfo{journal}{Applied energy} \bibinfo{volume}{184}
  (\bibinfo{year}{2016}) \bibinfo{pages}{987--994}.
\bibitem[{G{\"o}ransson et~al.(2014)G{\"o}ransson, Goop, Unger, Odenberger, and
  Johnsson}]{goransson2014linkages}
\bibinfo{author}{L.~G{\"o}ransson}, \bibinfo{author}{J.~Goop},
  \bibinfo{author}{T.~Unger}, \bibinfo{author}{M.~Odenberger},
  \bibinfo{author}{F.~Johnsson},
\newblock \bibinfo{title}{{Linkages between demand-side management and
  congestion in the European electricity transmission system}},
\newblock \bibinfo{journal}{Energy} \bibinfo{volume}{69} (\bibinfo{year}{2014})
  \bibinfo{pages}{860--872}.
\bibitem[{Henze et~al.(2004)Henze, Felsmann, and Knabe}]{henze2004evaluation}
\bibinfo{author}{G.~P. Henze}, \bibinfo{author}{C.~Felsmann},
  \bibinfo{author}{G.~Knabe},
\newblock \bibinfo{title}{{Evaluation of optimal control for active and passive
  building thermal storage}},
\newblock \bibinfo{journal}{International Journal of Thermal Sciences}
  \bibinfo{volume}{43} (\bibinfo{year}{2004}) \bibinfo{pages}{173--183}.
\bibitem[{Lizana et~al.(2017)Lizana, Chacartegui, Barrios-Padura, and
  Valverde}]{lizana2017advances}
\bibinfo{author}{J.~Lizana}, \bibinfo{author}{R.~Chacartegui},
  \bibinfo{author}{A.~Barrios-Padura}, \bibinfo{author}{J.~M. Valverde},
\newblock \bibinfo{title}{{Advances in thermal energy storage materials and
  their applications towards zero energy buildings: A critical review}},
\newblock \bibinfo{journal}{Applied Energy} \bibinfo{volume}{203}
  (\bibinfo{year}{2017}) \bibinfo{pages}{219--239}.
\bibitem[{Gils(2012)}]{gils2012gis}
\bibinfo{author}{H.~C. Gils},
\newblock \bibinfo{title}{{A GIS-based assessment of the district heating
  potential in Europe}},
\newblock \bibinfo{journal}{Deutsches Zentrum f{\"u}r Luft-und Raumfahrt (DLR):
  Graz, Austria}  (\bibinfo{year}{2012}).
\bibitem[{{Persson, U., M{\"o}ller, B., Wiechers, E.}(2017)}]{persson_2017}
\bibinfo{author}{{Persson, U., M{\"o}ller, B., Wiechers, E.}},
  \bibinfo{title}{{Heat Roadmap Europe. Deliverable 2.3: A final report
  outlining the methodology and assumptions used in the mapping}},
  \bibinfo{year}{2017}.
\bibitem[{Connolly et~al.(2014)Connolly, Lund, Mathiesen, Werner, M{\"o}ller,
  Persson, Boermans, Trier, {\O}stergaard, and Nielsen}]{connolly2014heat}
\bibinfo{author}{D.~Connolly}, \bibinfo{author}{H.~Lund},
  \bibinfo{author}{B.~V. Mathiesen}, \bibinfo{author}{S.~Werner},
  \bibinfo{author}{B.~M{\"o}ller}, \bibinfo{author}{U.~Persson},
  \bibinfo{author}{T.~Boermans}, \bibinfo{author}{D.~Trier},
  \bibinfo{author}{P.~A. {\O}stergaard}, \bibinfo{author}{S.~Nielsen},
\newblock \bibinfo{title}{{Heat Roadmap Europe: Combining district heating with
  heat savings to decarbonise the EU energy system}},
\newblock \bibinfo{journal}{Energy Policy} \bibinfo{volume}{65}
  (\bibinfo{year}{2014}) \bibinfo{pages}{475--489}.
\bibitem[{Schmid et~al.(2012)Schmid, Beer, and Corradini}]{corradini_2012}
\bibinfo{author}{T.~Schmid}, \bibinfo{author}{M.~Beer},
  \bibinfo{author}{R.~Corradini},
\newblock \bibinfo{title}{{Energiemodell der Wohngeb{\"a}ude}},
\newblock \bibinfo{journal}{BWK} \bibinfo{volume}{64} (\bibinfo{year}{2012})
  \bibinfo{pages}{{48--53}}.
\bibitem[{{Forschungsstelle f{\"u}r Energiewirtschaft}(2016)}]{ffe_merit_2016}
\bibinfo{author}{{Forschungsstelle f{\"u}r Energiewirtschaft}},
  \bibinfo{title}{{Merit Order der Energiespeicherung im Jahr 2030,
  Hauptbericht Teil 2: Techno{\"o}konomische Analyse Funktionaler
  Energiespeicher}}, \bibinfo{year}{2016}.
\bibitem[{{Gschwender, D., et al.}(2016)}]{lubw_2016_energieatlas}
\bibinfo{author}{{Gschwender, D., et al.}}, \bibinfo{title}{{Energieatlas
  Baden-W{\"u}rttemberg, Daten und Fakten zur Energiewende}},
  \bibinfo{year}{2016}.
\bibitem[{Gils(2015)}]{gils2015balancing}
\bibinfo{author}{H.~C. Gils}, \bibinfo{title}{{Balancing of intermittent
  renewable power generation by demand response and thermal energy storage}},
  Ph.D. thesis, {University of Stuttgart}, \bibinfo{year}{2015}.
\bibitem[{Grein and Pehnt(2011)}]{grein2011load}
\bibinfo{author}{A.~Grein}, \bibinfo{author}{M.~Pehnt},
\newblock \bibinfo{title}{{Load management for refrigeration systems:
  Potentials and barriers}},
\newblock \bibinfo{journal}{Energy Policy} \bibinfo{volume}{39}
  (\bibinfo{year}{2011}) \bibinfo{pages}{5598--5608}.
\bibitem[{{Schucht et al.}(2017)}]{nep2017nep}
\bibinfo{author}{{Schucht et al.}}, \bibinfo{title}{{Netzentwicklungsplan Strom
  2030, Version 2017 - Zweiter Entwurf der {\"U}bertragungsnetzbetreiber}},
  \bibinfo{year}{2017}.
\bibitem[{{N. Diefenbach, H. Cischinsky, M. Rodenfels, K.-D.
  Clausnitzer}(2010)}]{iwu2010_datenbasis}
\bibinfo{author}{{N. Diefenbach, H. Cischinsky, M. Rodenfels, K.-D.
  Clausnitzer}}, \bibinfo{title}{{Datenbasis Geb{\"a}udebestand}},
  \bibinfo{howpublished}{Institut Wohnen und Umwelt, Bremer Energie Institut},
  \bibinfo{year}{2010}.
\bibitem[{{T. Loga, B. Stein, N. Diefenbach, R.
  Born}(2015)}]{iwu2015wohngebaeudetypo}
\bibinfo{author}{{T. Loga, B. Stein, N. Diefenbach, R. Born}},
  \bibinfo{title}{{Deutsche Wohngeb{\"a}udetypologie}},
  \bibinfo{howpublished}{Institut Wohnen und Umwelt}, \bibinfo{year}{2015}.
\bibitem[{Medjroubi et~al.(2017)Medjroubi, M{\"u}ller, Scharf, Matke, and
  Kleinhans}]{medjroubi2017open}
\bibinfo{author}{W.~Medjroubi}, \bibinfo{author}{U.~P. M{\"u}ller},
  \bibinfo{author}{M.~Scharf}, \bibinfo{author}{C.~Matke},
  \bibinfo{author}{D.~Kleinhans},
\newblock \bibinfo{title}{{Open data in power grid modelling: new approaches
  towards transparent grid models}},
\newblock \bibinfo{journal}{Energy Reports} \bibinfo{volume}{3}
  (\bibinfo{year}{2017}) \bibinfo{pages}{14--21}.
\bibitem[{Morrison(2018)}]{morrison2018energy}
\bibinfo{author}{R.~Morrison},
\newblock \bibinfo{title}{{Energy system modeling: Public transparency,
  scientific reproducibility, and open development}},
\newblock \bibinfo{journal}{Energy strategy reviews} \bibinfo{volume}{20}
  (\bibinfo{year}{2018}) \bibinfo{pages}{49--63}.
\bibitem[{Pfenninger et~al.(2018)Pfenninger, Hirth, Schlecht, Schmid, Wiese,
  Brown, Davis, Gidden, Heinrichs, Heuberger et~al.}]{pfenninger2018opening}
\bibinfo{author}{S.~Pfenninger}, \bibinfo{author}{L.~Hirth},
  \bibinfo{author}{I.~Schlecht}, \bibinfo{author}{E.~Schmid},
  \bibinfo{author}{F.~Wiese}, \bibinfo{author}{T.~Brown},
  \bibinfo{author}{C.~Davis}, \bibinfo{author}{M.~Gidden},
  \bibinfo{author}{H.~Heinrichs}, \bibinfo{author}{C.~Heuberger}, et~al.,
\newblock \bibinfo{title}{{Opening the black box of energy modelling:
  Strategies and lessons learned}},
\newblock \bibinfo{journal}{Energy Strategy Reviews} \bibinfo{volume}{19}
  (\bibinfo{year}{2018}) \bibinfo{pages}{63--71}.
\bibitem[{{Federal Statistical Office of Germany}(2014)}]{destatis2014overview}
\bibinfo{author}{{Federal Statistical Office of Germany}},
  \bibinfo{title}{{2011 Census - Buildings and dwellings: Overview of
  variables, terms and related characteristics}}, \bibinfo{year}{2014}.
  \bibinfo{note}{\href{https://www.zensus2011.de/SharedDocs/Downloads/EN/Variables/Variables_buildings_and_housing.pdf?__blob=publicationFile&v=10}{https://www.zensus2011.de/SharedDocs/}}.
\bibitem[{{Research Data Centre of the Federal Statistical Office of Germany
  and the Statistical Offices of the
  L{\"a}nder}(2018)}]{census_sonderauswertung}
\bibinfo{author}{{Research Data Centre of the Federal Statistical Office of
  Germany and the Statistical Offices of the L{\"a}nder}},
  \bibinfo{title}{{Census 2011}}, \bibinfo{year}{2018}.
\bibitem[{H{\"o}hne(2015)}]{hoehne2015safe}
\bibinfo{author}{J.~H{\"o}hne},
\newblock \bibinfo{title}{{Das Geheimhaltungsverfahren SAFE}},
\newblock \bibinfo{journal}{Zeitschrift f{\"u}r amtliche Statistik Berlin
  Brandenburg} \bibinfo{volume}{2} (\bibinfo{year}{2015}).
\bibitem[{din(2010)}]{din2011_18599}
\bibinfo{title}{{DIN V 18599: 2010 - Energy efficiency of buildings-Calculation
  of the net, final and primary energy demand for heating, cooling,
  ventilation, domestic hot water and lighting}}, \bibinfo{year}{2010}.
\bibitem[{din(2008)}]{din2008_12831}
\bibinfo{title}{{DIN EN 12831: 2008 - Energy performance of buildings - Method
  for calculation of the design heat load}}, \bibinfo{year}{2008}.
\bibitem[{vdi(2008)}]{vdi2008_3807}
\bibinfo{title}{{VDI 3807: 2008 - Characteristic consumption values for
  buildings - Fundamentals}}, \bibinfo{year}{2008}.
\bibitem[{Bigalke et~al.(2015)Bigalke, Discher, Kunde, Schmidt, Zeng, Bensmann,
  and Stolte}]{bigalke2015dena_geb_rep}
\bibinfo{author}{U.~Bigalke}, \bibinfo{author}{H.~Discher},
  \bibinfo{author}{J.~Kunde}, \bibinfo{author}{M.~Schmidt},
  \bibinfo{author}{Y.~Zeng}, \bibinfo{author}{K.~Bensmann},
  \bibinfo{author}{C.~Stolte},
\newblock \bibinfo{title}{{Der dena-Geb{\"a}udereport 2015}},
\newblock \bibinfo{journal}{Deutsche Energie-Agentur (dena). Berlin}
  (\bibinfo{year}{2015}).
\bibitem[{Glombik(2008)}]{glombik2008energieeffiziente}
\bibinfo{author}{B.~M. Glombik}, \bibinfo{title}{{Energieeffiziente
  Bestandsentwicklung - Die Umsetzung von Energiestandards in den Best{\"a}nden
  der Wohnungswirtschaft}}, \bibinfo{howpublished}{RWTH Aachen},
  \bibinfo{year}{2008}.
\bibitem[{Holtfort(2002)}]{holtfort2002enev}
\bibinfo{author}{J.~Holtfort},
\newblock \bibinfo{title}{{In-Kraft-Treten der EnEV 2002}},
\newblock \bibinfo{journal}{IKZ-Haustechnik} \bibinfo{volume}{3}
  (\bibinfo{year}{2002}).
\bibitem[{{Bundesministerium f{\"u}r Wirtschaft und
  Energie}(2018)}]{bmwi2018energiedaten}
\bibinfo{author}{{Bundesministerium f{\"u}r Wirtschaft und Energie}},
  \bibinfo{title}{{Zahlen und Fakten Energiedaten}}, \bibinfo{year}{2018}.
\bibitem[{{Statistische {\"A}mter des Bundes und der
  L{\"a}nder}(2013)}]{destatis2011census}
\bibinfo{author}{{Statistische {\"A}mter des Bundes und der L{\"a}nder}},
  \bibinfo{title}{{Zenus 2011}}, \bibinfo{year}{2013}.
  \bibinfo{note}{\href{https://www.zensus2011.de/DE/Home/home_node.html}{https://www.zensus2011.de}}.
\bibitem[{{Nabe, C.}(2011)}]{Nabe_Prognos_2011}
\bibinfo{author}{{Nabe, C.}}, \bibinfo{title}{{Potenziale der W\"armepumpe zum
  Lastmanagement im Strommarkt und zur Netzintegration erneuerbarer Energien}},
  \bibinfo{howpublished}{{Prognos AG, Ecofys Germany GmbH}},
  \bibinfo{year}{2011}.
\bibitem[{{Lange, M., Zobel, M.}(2017)}]{lange_2018_novaref}
\bibinfo{author}{{Lange, M., Zobel, M.}},
  \bibinfo{title}{{\href{https://www.tib.eu/de/suchen/id/TIBKAT\%3A894155547/Schlussbericht-zum-Vorhaben-Erstellung-neuer-Referenzlastprofile/}{Schlussbericht
  zum Vorhaben: Erstellung neuer Referenzlastprofile zur Auslegung,
  Dimensionierung und Wirtschaftlichkeitsberechnung von
  Hausenergieversorgungssystemen (NOVAREF)}}},
  \bibinfo{howpublished}{{DLR-Institut f\"ur Vernetzte Energiesysteme,
  F{\"o}rderkennzeichen BMWi 03FS14013}}, \bibinfo{year}{2017}.
\bibitem[{{Kiesel, R{\"u}diger}(2016)}]{kiesel_2016_modelling}
\bibinfo{author}{{Kiesel, R{\"u}diger}}, \bibinfo{title}{{Modelling Day-Ahead
  and Intraday Electricity Markets}}, \bibinfo{howpublished}{Chair for Energy
  Trading and Finance, University of Duisburg-Essen}, \bibinfo{year}{2016}.
\bibitem[{Allen(1976)}]{allen1976modified}
\bibinfo{author}{J.~Allen},
\newblock \bibinfo{title}{{A Modified Sine Wave Method for Calculating Degree
  Days}},
\newblock \bibinfo{journal}{Environmental Entomology} \bibinfo{volume}{5}
  (\bibinfo{year}{1976}).
\bibitem[{{Institut f\"ur Wohnen und Umwelt}(2018)}]{iwu_exel_tool}
\bibinfo{author}{{Institut f\"ur Wohnen und Umwelt}}, \bibinfo{title}{{IWU
  Excel Tool - Gradtagszahlen Deutschland}}, \bibinfo{year}{2018}.
  \bibinfo{note}{\href{https://www.iwu.de/fileadmin/user_upload/dateien/energie/werkzeuge/Gradtagszahlen_Deutschland.xls}{www.iwu.de}}.
\bibitem[{{Statistisches Bundesamt}(2012)}]{destatis_microcen_2010}
\bibinfo{author}{{Statistisches Bundesamt}}, \bibinfo{title}{{Mikrozensus -
  Zusatzerhebung 2010 Bestand und Struktur der Wohneinheiten Wohnsituation der
  Haushalte}}, \bibinfo{year}{2012}.
\bibitem[{Christidis et~al.(2017)Christidis, Mollenhauer, Tsatsaronis,
  Schuchardt, Holler, B{\"o}ttger, and Bruckner}]{christidis_2017_eneff}
\bibinfo{author}{A.~Christidis}, \bibinfo{author}{E.~Mollenhauer},
  \bibinfo{author}{G.~Tsatsaronis}, \bibinfo{author}{G.~Schuchardt},
  \bibinfo{author}{S.~Holler}, \bibinfo{author}{D.~B{\"o}ttger},
  \bibinfo{author}{T.~Bruckner},
\newblock \bibinfo{title}{{EnEff-W{\"a}rme: Einsatz von W{\"a}rmespeichern und
  Power-to-Heat-Anlagen in der Fernw{\"a}rmeerzeugung}},
\newblock \bibinfo{journal}{AGFW, Frankfurt am Main, Germany}
  (\bibinfo{year}{2017}).
\bibitem[{Stadler(2008)}]{stadler2008gigantisches}
\bibinfo{author}{I.~Stadler},
\newblock \bibinfo{title}{{Ein gigantisches Speicherpotenzial}},
\newblock \bibinfo{journal}{Solarzeitalter} \bibinfo{volume}{1}
  (\bibinfo{year}{2008}) \bibinfo{pages}{60--64}.
\bibitem[{Gassmann(2018)}]{gassmann2018unerwartete}
\bibinfo{author}{M.~Gassmann},
\newblock \bibinfo{title}{{Das unerwartete Comeback des Nachtspeicherofens}},
\newblock \bibinfo{journal}{{Die Welt}}  (\bibinfo{year}{2018}).
\bibitem[{{Fraunhofer IWES/IFAM, Stiftung
  Umweltenergierecht}(2014)}]{energiewende2014power}
\bibinfo{author}{{Fraunhofer IWES/IFAM, Stiftung Umweltenergierecht}},
  \bibinfo{title}{{Power-to-Heat zur Integration von ansonsten abgeregeltem
  Strom aus Erneuerbaren Energien, Studie im Auftrag der Agora Energiewende}},
  \bibinfo{year}{2014}.
\bibitem[{{OECD Directorate for Public Governance and Territorial
  Development}(2011)}]{oecd2011regional}
\bibinfo{author}{{OECD Directorate for Public Governance and Territorial
  Development}}, \bibinfo{title}{{OECD Regional Typology}},
  \bibinfo{year}{2011}.
\bibitem[{Shlomo(2011)}]{shlomo2011unterschiede}
\bibinfo{author}{J.~B. Shlomo}, \bibinfo{title}{{Unterschiede in den
  Eigentumsquoten von Wohnimmobilien: Erkl{\"a}rungsversuche und
  Wirkungsanalyse}}, \bibinfo{publisher}{WHL}, \bibinfo{year}{2011}.
\bibitem[{Bach et~al.(2016)Bach, Werling, Ommen, M{\"u}nster, Morales, and
  Elmegaard}]{bach2016integration}
\bibinfo{author}{B.~Bach}, \bibinfo{author}{J.~Werling},
  \bibinfo{author}{T.~Ommen}, \bibinfo{author}{M.~M{\"u}nster},
  \bibinfo{author}{J.~M. Morales}, \bibinfo{author}{B.~Elmegaard},
\newblock \bibinfo{title}{{Integration of large-scale heat pumps in the
  district heating systems of Greater Copenhagen}},
\newblock \bibinfo{journal}{Energy} \bibinfo{volume}{107}
  (\bibinfo{year}{2016}) \bibinfo{pages}{321--334}.
\bibitem[{{Plazzo, Michaela}(2017)}]{plazzo_2017_P2H_berlin}
\bibinfo{author}{{Plazzo, Michaela}}, \bibinfo{title}{{Vattenfall
  Power-to-Heat-Anlage in Spandau soll Fernw{\"a}rme f{\"u}r bis zu 30.000
  Haushalte liefern}}, \bibinfo{howpublished}{EUWID Neue Energie,
  \href{https://www.euwid-energie.de/vattenfall-power-to-heat-anlage-in-spandau-soll-fernwaerme-fuer-bis-zu-30-000-haushalte-liefern/}{https://www.euwid-energie.de/}},
  \bibinfo{year}{2017}.
\bibitem[{{Kiefert, Ulrike}(2018)}]{kiefert_2018_power}
\bibinfo{author}{{Kiefert, Ulrike}}, \bibinfo{title}{{Power-to-Heat-Anlage
  bekommt drei Kessel}}, \bibinfo{howpublished}{Spandauer Volksblatt,
  \href{https://www.berliner-woche.de/siemensstadt/c-wirtschaft/power-to-heat-anlage-bekommt-drei-kessel_a170234}{https://www.berliner-woche.de}},
  \bibinfo{year}{2018}.
\bibitem[{{Prognos, EWI, GWS}(2014)}]{prognos_entwicklung_2014}
\bibinfo{author}{{Prognos, EWI, GWS}}, \bibinfo{title}{{Entwicklung der
  Energiem{\"a}rkte - Energiereferenzprognose}}, \bibinfo{year}{2014}.
\bibitem[{{Beer, M., et al.}(2012)}]{ffe_2012_flex}
\bibinfo{author}{{Beer, M., et al.}}, \bibinfo{title}{{Flex - Flexible
  Betriebsweise von Kraft-W{\"a}rme-Kopplungsanlagen}}, \bibinfo{year}{2012}.
\bibitem[{Wei et~al.(2014)Wei, Jones, and De~Wilde}]{wei2014driving}
\bibinfo{author}{S.~Wei}, \bibinfo{author}{R.~Jones},
  \bibinfo{author}{P.~De~Wilde},
\newblock \bibinfo{title}{{Driving factors for occupant-controlled space
  heating in residential buildings}},
\newblock \bibinfo{journal}{Energy and Buildings} \bibinfo{volume}{70}
  (\bibinfo{year}{2014}) \bibinfo{pages}{36--44}.
\bibitem[{Lee~Rodgers and Nicewander(1988)}]{lee1988thirteen}
\bibinfo{author}{J.~Lee~Rodgers}, \bibinfo{author}{W.~A. Nicewander},
\newblock \bibinfo{title}{{Thirteen ways to look at the correlation
  coefficient}},
\newblock \bibinfo{journal}{The American Statistician} \bibinfo{volume}{42}
  (\bibinfo{year}{1988}) \bibinfo{pages}{59--66}.
\bibitem[{{Bundesverband W{\"a}rmepumpe}(2016)}]{bwp_waermepump_2016}
\bibinfo{author}{{Bundesverband W{\"a}rmepumpe}},
  \bibinfo{title}{{W{\"a}rmepumpen Bestand in Deutschland}},
  \bibinfo{year}{2016}.

\end{thebibliography}







\end{document}